\documentclass[aps,pre,showpacs,groupedaddress,reprint,floatfix]{revtex4-1}

\usepackage{graphicx}
\usepackage{dcolumn}
\usepackage{tabularx}
\usepackage{amsfonts}
\usepackage{amssymb}
\usepackage{mathtools}
\usepackage{chemarrow}
\usepackage[pdfencoding=auto]{hyperref}
\usepackage{dsfont}
\usepackage{psfrag}

\usepackage{color}
\newcommand{\review}[1]{\textcolor{black}{#1}}
\newcommand{\eqreview}[1]{\color{black}{#1}}

\usepackage[normalem]{ulem}

\newcommand{\bfx}{{\mathbf x}}
\newcommand{\ud}{\mathrm{d}}
\newcommand{\bb}{\mathsf{b}}
\newcommand{\rr}{\mathsf{r}}

\begin{document}

\title{Reactions, Diffusion and Volume Exclusion in a Heterogeneous System of Interacting Particles}

\author{Daniel B. Wilson}
 \email{daniel.wilson@maths.ox.ac.uk}
\author{Helen Byrne}%
\email{helen.byrne@maths.ox.ac.uk}
\author{Maria Bruna}
\email{bruna@maths.ox.ac.uk}
\affiliation{%
 Mathematical Institute, University of Oxford, Radcliffe Observatory Quarter, Woodstock Road, Oxford, OX2 6GG, United Kingdom
}%
\date{\today}

\begin{abstract}
\review{Complex biological and physical transport processes are often described through systems of interacting particles. Excluded-volume effects on these transport processes are well studied, however the interplay between volume exclusion and reactions between heterogenous particles is less well known. In this paper we develop a novel framework for modeling reaction-diffusion processes which directly incorporates volume exclusion. From an off-lattice microscopic individual based model we use the Fokker--Planck equation and the method of matched asymptotic expansions to derive a low-dimensional macroscopic system of nonlinear partial differential equations describing the evolution of the particles. A biologically motivated, hybrid model of chemotaxis with volume exclusion is explored, where reactions occur at rates dependent upon the chemotactic environment. Further,} we show that for reactions due to contact interactions the appropriate reaction term in the macroscopic model is of lower order in the asymptotic expansion than the nonlinear diffusion term. However, we find that the next reaction term in the expansion is needed to ensure good agreement with simulations of the microscopic model. \review{Our macroscopic model allows for more direct parameterization to experimental data than the models available to date.}
\end{abstract}

\pacs{{05.10.Gg},{ 05.40.Fb},{ 02.30.Jr},{ 02.30.Mv}}

\maketitle

\section{Introduction}
Cellular migration plays an important role in many biological processes, including tumor growth and invasion through an extracellular matrix \cite{Chauviere2010,GoOrGrow}, the formation of blood vessels via the movement of endothelial cells during embryogenesis \cite{Drake2003,angio_nature1997} and the directed motion of immune cells to infected sites \cite{robertson1993,paracrine2004}. \review{Classical continuum models assume that cells move down spatial gradients in their density (diffusion) and/or up spatial gradients in the concentration of a diffusible chemical (chemotaxis). However, these models treat the cells as point particles. In recent years experimental evidence has revealed the importance of excluded-volume effects in a variety of biological applications such as cellular migration \cite{mort2016reconciling,kim2010crowding,gejji2012macroscopic,plank2012models} and molecular traffic \cite{neri2011totally,graf2017generic,chowdhury2005physics}, as well as physical processes including vehicular traffic \cite{chowdhury2000statistical} and queueing \cite{kipnis1986central}. These findings have motivated a significant amount of theoretical research focussed on incorporating excluded-volume effects into mathematical models of transport \cite{sun2007toward,lushnikov2008macroscopic,Pillay:2018JMB}.} 

One approach is to discretize space into a regular lattice, and allow at most one \review{particle} to occupy each lattice site \review{\cite{Simpson:2009gi,ascolani2013exclusion}}. \review{Diffusion is represented by jumps between neighboring lattices and these are aborted if the neighbouring lattice site is already occupied.} \review{An alternative lattice-based approach which incorporates finite-size effects is the cellular Potts model \cite{lushnikov2008macroscopic}.}  An advantage of \review{these} lattice-based \review{approaches} is that it is relatively straightforward to include complex  individual-based mechanisms, and to obtain a macroscopic model based on partial differential equations (PDEs). However, the form of the lattice can introduce artefacts into these models, such as a bias towards homogeneity \cite{artefacts1, artefacts2}.
 
An alternative approach is to derive continuum equations from an off-lattice individual-based model, starting from Brownian particles with excluded-volume interactions. \review{For a single species of particles one obtains a nonlinear diffusion equation \cite{bruna2012single}, whereas on-lattice models yield a linear diffusion equation in the continuum limit \cite{Simpson:2009gi}. An off-lattice model with volume exclusion, but with motility rules inherited from the standard lattice-based approach is considered in \cite{Dyson12}. In this case, the resulting PDE model is also a nonlinear diffusion equation similar to \cite{bruna2012single}.}

As in the case of diffusion only, stochastic reaction-diffusion processes have been modeled using on- and off-lattice approaches \review{\cite{meinecke2016excluded,seki2012diffusion}}. A challenge with lattice-based approaches is that it is difficult to combine reactions and finite-size interactions. As mentioned above, in the latter case the lattice size is chosen so that at most  one cell occupies a single lattice site. By contrast, the usual lattice-based approach for bimolecular reactions requires two cells to be in the same compartment in order to react \cite{erban2009stochastic}. 
The standard off-lattice approach to model reaction-diffusion processes is based on the Smoluchowski model, which represents cells as point particles undergoing Brownian motion and reacting whenever they become closer than a given reaction radius \cite{andrews2004stochastic}. The reaction radius is chosen as a function of the reaction rate and diffusion constants of the reactants, and is typically independent of their physical radii. In fact, it has been shown that the reaction radius may be unrealistically smaller than the radii of the reactants \cite{erban2009stochastic}. Several methods have been proposed to overcome the shortcomings of these models \cite{andrews2004stochastic,isaacson2013convergent}, including one where the reaction radius can be fixed to represent the reactant radius \cite{erban2009stochastic}, but to our knowledge none includes the excluded-volume effects \review{which have been shown to significantly impact diffusion-limited reactions \cite{park2003excluded}}. To this end, in this paper we \review{extend the off-lattice} model in \cite{bruna2012single} \review{to account} for reactions between two subpopulations of hard-sphere Brownian particles. The advantage of our approach to modelling reactions is that the sizes of the \review{particles} (and therefore excluded-volume interactions) are included explicitly in our individual-based model, thereby circumventing the problem of unrealistically small reaction radii and providing a more natural way to describe diffusion-limited reactions \cite{erban2009stochastic}. 

\review{The remainder of the paper is organised as follows. In Section \ref{Spont} we introduce unimolecular reactions to represent spontaneous switching between subpopulations. These simple reactions allow us to develop our framework, but can also be motivated by the phenotypic switching of tumor cells within a glioblastoma \cite{GoOrGrow} as well as other canonical studies in mathematical biology such as infectious diseases and social interaction networks \cite{MurrayBook1}. We first formulate the discrete model for the interacting particles and then use the Fokker--Planck equation in combination with the method of matched asymptotic expansions to derive a low-dimensional system of nonlinear reaction-diffusion PDEs. A numerical example comparing the microscopic and macroscopic models is presented in Subsection \ref{CollNR}. In Subsection \ref{chemo} we apply our framework to cellular chemotaxis, by making the rate at which cells switch between phenotypes depend on the concentration of a third species which also acts as a chemotattractant for one cell population. In Section \ref{Coll} we study the case when reactions between subpopulations (bimolecular reactions) occur in response to collisions. A biological motivation for switching due to local interactions can be found in juxtacrine signaling as illustrated by the delta-notch signaling pathway \cite{Notch}. }
Finally in Section \ref{Discuss} we summarize our results and suggest possible directions for future investigations.

\section{Spontaneous Switching: Unimolecular Reactions} \label{Spont}

\review{We} place our work \review{in} context \review{by briefly summarizing} the results of  \cite{bruna2012multiple}, which considers two subpopulations of inert Brownian particles with hard-core excluded-volume interactions. Consider a population of $N$ hard-spheres of diameter $\epsilon \ll 1$ within a bounded domain $\Omega \subset \mathbb{R}^d$ \review{of typical dimensionless volume of order one}, where $d =2,3$. There are two subpopulations of spheres, which we term ``red" and ``blue". Let there be $N_{\bb}$ blue spheres and $N_{\rr}$ red spheres, where $N=N_{\bb}+N_{\rr}$. \review{Assume that the particles occupy a small volume fraction, so that $(N_{\bb} + N_{\rr}) \epsilon ^d \ll 1$. Let $\mathcal I_{\bb}$ denote the set of indices corresponding to blue particles, and $\mathcal I_{\rr}$ the indices corresponding to red particles.} The blue and red spheres have diffusion coefficients $D_{\bb}$ and $D_{\rr}$, and drift vectors ${\mathbf f}_\bb$ and ${\mathbf f}_\rr$, respectively. The center of the $i$th particle is denoted by ${\mathbf X}_i(t)$ and evolves according to the overdamped Langevin SDE:
\begin{subequations} \label{model_bruna}
\begin{alignat}{2}
\mathrm{d}{\mathbf X}_i(t) &= \sqrt{2D_{\bb}} \mathrm{d}{\mathbf W}_i(t) + {\mathbf f}_\bb({\mathbf X}_i(t)) \mathrm{d}t, & \ \ & \eqreview{i \in \mathcal I_{\bb},}  \label{2pop_SDEb} \\
\mathrm{d}{\mathbf X}_i(t) &= \sqrt{2D_{\rr}} \mathrm{d}{\mathbf W}_i(t) + {\mathbf f}_\rr({\mathbf X}_i(t)) \mathrm{d}t,  && \eqreview{i \in \mathcal I_{\rr},} \label{2pop_SDEr}
\end{alignat}
\end{subequations}
\review{for $1 \leq i \leq N$,} where the ${\mathbf W}_i$ are independent, $d$-dimensional standard Brownian motions. Reflective boundary conditions are imposed whenever two spheres are in contact \review{($\| {\bf X}_i-{\bf X}_j\| = \epsilon$ for $i \ne j$)} and also on the boundary of the domain $\partial \Omega$. From a particle-level description, the method of matched asymptotic expansions for a small but finite volume fraction is used to obtain a continuum model for the marginal probability density functions, $b({\mathbf x},t)$ and $r({\mathbf x},t)$ for blue and red spheres, respectively. The continuum model comprises the following system of nonlinear cross-diffusion equations
\renewcommand*{\arraystretch}{.89}
\begin{subequations} \label{2pop_p}
\begin{align}
\label{2pop_pPDEs}
\partial_t \begin{pmatrix} b \\ r \end{pmatrix} &=  \nabla_{{\mathbf x}} \cdot \left[ {\mathcal D} \,  \nabla_{{\mathbf x}} \! \begin{pmatrix} b \\ r \end{pmatrix} - {\mathcal F}\begin{pmatrix} b \\ r \end{pmatrix}
 \right],
\end{align}
\review{for ${\mathbf x} \in \Omega \subset \mathbb{R}^d$, }with zero-flux boundary conditions on $\partial \Omega$. In equations (\ref{2pop_pPDEs}), the nonlinear diffusion \review{$\mathcal{D}$ and drift ${\mathcal F}$ matrices are defined as follows:}
\begin{align}\label{diff_matrix_no_switching}
 {\mathcal D} = \text{diag}(D_{\bb}, D_{\rr}) + \epsilon^d \widetilde{\mathcal D}(b,r), 
\end{align}
where
\begin{align*}
\widetilde{\mathcal D} = 
\begin{pmatrix} D_{\bb} (N_{\bb}-1) \omega b    -  D_{\bb} N_{\rr} \eta_{\bb} r  \hspace{1cm} D_{\bb} N_{\rr} \xi_{\bb} b   \\ \\ D_{\rr} N_{\bb}  \xi_{\rr} r \hspace{1cm}  D_{\rr} (N_{\rr}-1) \omega r   -  D_{\rr} N_{\bb} \eta_{\rr} b \end{pmatrix} ,
 \end{align*}
 and
\begin{align}\label{drift_matrix_no_switching}
 {\mathcal F} = \begin{pmatrix} {\mathbf f}_\bb & N_{\rr} \epsilon^d \eta_{\bb} ( {\mathbf f}_\rr - {\mathbf f}_\bb ) b \\ N_{\bb} \epsilon^d \eta_{\rr} ( {\mathbf f}_\bb - {\mathbf f}_\rr ) r & {\mathbf f}_\rr \end{pmatrix}  ,
 \end{align}
 \end{subequations}
 where the constants \review{$ \omega, \xi_k, \eta_k$ (for $k = \bb, l = \rr$ and vice versa)} are defined by
 \begin{align}\label{parameters}
 \begin{aligned}
 \omega &= \frac{2\pi(d-1)}{d}, \quad 
 \xi_k = \dfrac{2\pi}{d}\dfrac{[(d-1)D_k + dD_l]}{D_k+D_l}, \\ \eta_k &= \dfrac{2\pi}{d}\dfrac{D_k}{D_k+D_l}.
  \end{aligned}
\end{align}  
The density-dependent terms in \eqref{diff_matrix_no_switching}, contained in $\widetilde {\mathcal D}$, reveal how collective diffusion changes due to self-crowding and competition between subpopulations.

In what follows we  extend model \eqref{2pop_p} to incorporate reactions between subpopulations. We introduce switching at the discrete level by allowing particles to spontaneously switch between the blue and red types at a constant rate that is independent of both time and space.

\subsection{Particle-based model} \label{Particle-based}

We consider again a population of $N$ spheres that can be either tagged blue or red, all of diameter $\epsilon$ in $\Omega \subset \mathbb R^d$, $d= 2,3$, bounded and of dimensionless volume of order one. \review{As before, we denote the position of the $i$th individual as ${\mathbf X}_i(t)$, and $\mathcal I_{\bb}(t)$ and $\mathcal I_{\rr}(t)$ the sets of indices corresponding to blue and red particles, respectively. However, we note that now these sets now depend on time since particles can change color. We denote by $S_i(t)$ the color of the $i$th individual at time $t$, where $S_i(t) = \bb$  if the particle is blue or $S_i(t) = \rr$ if it is red.} As before particles evolve according to overdamped Langevin SDEs:
\begin{subequations} \label{sde_switch}
\begin{alignat}{2}
\mathrm{d}{\mathbf X}_i(t) &= \sqrt{2D_{\bb}} \mathrm{d} {\mathbf W}_i(t) + {\mathbf f}_\bb({\mathbf X}_i(t)) \mathrm{d}t, & \ \ & \eqreview{i \in \mathcal I_{\bb}(t),} \label{SDE_SSb} \\
\mathrm{d} {\mathbf X}_i(t) &= \sqrt{2D_{\rr}} \mathrm{d}{\mathbf W}_i(t) + {\bf f}_\rr({\bf X}_i(t)) \mathrm{d}t, &  & \eqreview{i \in \mathcal I_{\rr}(t),}  \label{SDE_SSr}
\end{alignat}
for $1 \leq i \leq N$. We also introduce two unimolecular reactions to account for spontaneous switching:
\begin{equation} \label{reaction_spon}
B \xrightarrow{k_\bb} R, \qquad R \xrightarrow{k_\rr} B, 
\end{equation}
\end{subequations}
where $k_\bb$ and $k_\rr$ are the switching rates. When considering the probabilistic
description of the discrete model, the color of the $i$th particle is a random variable. 
We therefore consider the joint probability density function $P(\vec{x},\vec{s},t)$ \review{$= \mathbb{P}(\{ ({\bf X}_1(t) , \ldots , {\bf X}_N(t))  = \vec{x} \} \cap \{ (S_1(t) , \ldots , S_N(t))  = \vec{s} \} )$}, the
probability of the $N$ particles being in the configuration 
$\vec{x} = ({\bf x}_1,\ldots,{\bf x}_N)$ with the colors of the particles being 
described by the vector of states $\vec{s} = (s_1,\ldots,s_N)$ at time $t$. 
The configuration space (set of legal configurations) is defined as 
$\Omega_{\epsilon}^N \times \{\bb,\rr\}^N$, where $\Omega_{\epsilon}^N = \Omega^N\setminus \mathcal{B}_{\epsilon}$, and $\mathcal{B}_{\epsilon}$ is the 
  set of illegal spatial configurations,
\begin{equation*}
\mathcal{B}_{\epsilon} = \left\lbrace \vec{x} \in \Omega^N : \exists i \neq j \text{ such that } \|{\bf x}_i - {\bf x}_j \| \leq \epsilon \right\rbrace .
\end{equation*}

Using the Chapman--Kolmogorov equation \cite{RossIntroProb} we can write the following \review{Fokker--Planck equation as an exact description of} the evolution of the density  $P(\vec{x},\vec{s},t)$ in configuration space $\Omega_{\epsilon}^N \times \{\bb,\rr\}^N$,
\begin{subequations}
\begin{align} 
\label{SS_PPDE}
\begin{aligned}
	\partial_t P = \ & \nabla_{\vec{x}} \cdot \left[ {\mathsf D}_{\vec s} \nabla_{\vec{x}} P(\vec{x}, \vec{s},t) - \vec{F}(\vec x, \vec s) P(\vec{x}, \vec{s},t) \right] \\
	& + k_\bb \sum_{\vec{c} \in \mathcal{B}(\vec{s})} P(\vec{x},\vec{c},t) + k_\rr \sum_{\vec{c} \in \mathcal{R}(\vec{s})} P(\vec{x},\vec{c},t) \\
	& - \left( k_\bb |\mathcal{R}(\vec{s})| + k_\rr |\mathcal{B}(\vec{s})| \right)P(\vec{x}, \vec{s},t),
\end{aligned}	
\end{align}
where ${\mathsf D}_{\vec s} = \text{diag}(D_{s_1},\ldots,D_{s_N})$ and $\vec{F}(\vec{x}, \vec s) = ({\bf f}_{s_1}({\bf x}_1) , \ldots , {\bf f}_{s_N}({\bf x}_N) )$. \review{We denote by $\mathcal{C}^{N}$ the set of all possible color state vectors $\vec s$, and by $\mathcal{B}(\vec{s}), \mathcal{R}(\vec{s})   \subset \mathcal C^N$ the sets of state vectors that differ by one entry from $\vec s$: $\vec c \in \mathcal{B}(\vec{s})$ if it requires one blue particle to switch to red to become $\vec s$, and vice versa for those in $\mathcal{R}(\vec{s})$. For example, suppose that $N = 3$ and $\vec s = (\bb, \rr, \rr)$. Then $\mathcal B(\vec s) = \{ (\bb, \bb, \rr), (\bb, \rr, \bb) \}$ and $\mathcal R(\vec s) = \{ (\rr, \rr, \rr)\}$.} 
On the boundaries of the configuration space, $\partial \Omega_{\epsilon}^N$, we impose zero-flux boundary conditions
\begin{equation}
\left[ {\mathsf D}_{\vec{s}} \nabla_{\vec{x}} P(\vec{x}, \vec{s},t) - \vec{F}(\vec{x}, \vec s) P(\vec{x}, \vec{s},t) \right] \cdot \vec{n} = 0, \tag{5b} \label{SS_PPDE_BC}
\end{equation}
for all $\vec{s} \in \mathcal{C}^{N}$. Here $\vec{n} \in \mathcal{S}^{dN-1}$ denotes the unit outward normal. The initial condition is $P(\vec{x},\vec{s},0) = P_0(\vec{x},\vec{s})$, \review{with $P_0$ invariant to permutations of the labels of particles with the same color.} 
\end{subequations}

\subsection{Population-based model}

Although linear, the PDE for the particle-based model (\ref{SS_PPDE}) is very high-dimensional for large $N$. For this reason, we seek to derive a coupled system of low dimensional PDEs for the probability distribution of the position and state of a typical particle. Accordingly we introduce
\begin{subequations} \label{br_densities}
\begin{align}
b({\bf x},t) = \! \int_{\Omega_{\epsilon}^N}  \sum_{\vec{s} \in \mathcal{C}^{N}} P(\vec{x},\vec{s},t)\delta(s_1-\bb) \delta({\bf x}-{\bf x}_1) \mathrm{d}\vec{x},  \label{SS_blue_pdf}\\
r({\bf x},t) = \!\int_{\Omega_{\epsilon}^N} \sum_{\vec{s} \in \mathcal{C}^{N}} P(\vec{x},\vec{s},t) \delta(s_1-\rr)\delta({\bf x}-{\bf x}_1) \mathrm{d}\vec{x}. \label{SS_red_pdf}
\end{align}
\end{subequations}
\review{The particle choice is unimportant since $P$ is invariant with respect to permutations of particle position labels.}  We remark that the functions $b({\bf x},t)$ and $r({\bf x},t)$ are not marginal distribution functions as in \cite{bruna2012multiple}. Here they represent the probabilities of finding a particle at position ${\bf x}$ at time $t$, and the particle being either blue or red, respectively. We denote by $N_{\bb}(t)$ and $N_{\rr}(t)$ the number of blue and red particles at time $t$ and note from equations \eqref{br_densities} that $\int_{\Omega} b({\bf x},t) \mathrm{d}{\bf x} = N_{\bb}(t)/N$ and $\int_{\Omega} r({\bf x},t) \mathrm{d}{\bf x} = N_{\rr}(t)/N$. We shall refer to $b({\bf x},t)$ and $r({\bf x},t)$ as the blue and red density functions. 

As before, we focus on the low volume fraction regime \review{$N \epsilon^d \ll 1$} in which pairwise interactions dominate those involving three or more particles. Under these assumptions the problem reduces to the case when $N=2$ \cite{bruna2012multiple}. We introduce the following notation, $P_{s_1s_2}({\bf x}_1,{\bf x}_2,t) \equiv P({\bf x}_1,{\bf x}_2,s_1,s_2,t)$. For simplicity, we describe the derivation for ${\bf f}_\bb = {\bf f}_\rr = {\bf 0}$, but the case with drift can be obtained in a similar manner. Setting $N=2$ in equation (\ref{SS_PPDE}) we find
\begin{subequations}\label{FP_N2}
\begin{align}
\partial_t P_{\bb \bb} &=  \nabla_{\vec{x}} \!\cdot \! ( {\mathsf D}_{\bb \bb} \nabla_{\vec{x}} P_{\bb \bb})  + k_\rr ( P_{\bb \rr} + P_{\rr \bb} ) - 2 k_\bb P_{\bb \bb},  \label{SS_PPDE_bb}\\
\partial_t P_{\bb \rr} &=  \nabla_{\vec{x}} \!\cdot \! (  {\mathsf D}_{\bb \rr} \nabla_{\vec{x}} P_{\bb \rr})+  k_\bb P_{\bb \bb}  + k_\rr  P_{\rr \rr}   - ( k_\bb \!+\! k_\rr ) P_{\bb \rr},  \label{SS_PPDE_br}\\
\partial_t P_{\rr \bb} &=  \nabla_{\vec{x}} \!\cdot \! ( {\mathsf D}_{\rr \bb} \nabla_{\vec{x}} P_{\rr \bb} ) +  k_\bb P_{\bb \bb} + k_\rr  P_{\rr \rr}  - ( k_\bb \!+ \! k_\rr ) P_{\rr \bb},  \label{SS_PPDE_rb}\\
\partial_t P_{\rr \rr} &=  \nabla_{\vec{x}} \!\cdot \! ( {\mathsf D}_{\rr \rr} \nabla_{\vec{x}} P_{\rr \rr} ) + k_\bb ( P_{\bb \rr} + P_{\rr \bb} ) - 2 k_\rr P_{\rr \rr},  \label{SS_PPDE_rr}
\end{align}
\end{subequations} 
\review{where $\vec x = (\bfx_1, \bfx_2)$, with no-flux boundary conditions as in \eqref{SS_PPDE_BC}. The initial conditions on $P_{s_1s_2}$ follow from the initial condition on $P$. For example, $P_{\bb \bb}(\bfx_1, \bfx_2, 0) = P_0(\bfx_1, \bfx_2, \bb, \bb)$.} \review{For $N=2$, equations \eqref{br_densities} become } 
\begin{subequations}
\begin{align}
b({\bf x}_1,t) = \int_{\Omega_{\epsilon}({\bf x}_1)} ( P_{\bb \bb} + P_{\bb \rr}) \,\mathrm{d}{\bf x}_2, \label{SS_blue_pdf2} \\
r({\bf x}_1,t) = \int_{\Omega_{\epsilon}({\bf x}_1)} ( P_{\rr \rr}+ P_{\rr \bb}) \, \mathrm{d}{\bf x}_2,  \label{SS_red_pdf2}
\end{align}
\review{where $\Omega_{\epsilon}({\bf x}_1)$ denotes the region available to the second particle when the first particle is at $\bfx_1$, namely, $\Omega_\epsilon (\bfx_1) = \Omega \setminus B_\epsilon(\bfx_1)$. Since the domain dimensions are much larger than the particles' diameter, the volume $|\Omega_\epsilon (\bfx_1)|$ is constant to leading order.}
\end{subequations}
Combining equations (\ref{SS_PPDE_bb}), (\ref{SS_PPDE_br}) and (\ref{SS_blue_pdf2}) we arrive at the following equation for $b({\bf x}_1,t)$
\begin{align} \label{SS_blue_CI_1}
\begin{aligned}
	\partial_t b  = & \int_{\Omega_{\epsilon}({\bf x}_1)}  \nabla_{{\bf x}_1} \cdot \left( D_{\bb} \nabla_{{\bf x}_1} P_{\bb \bb} + D_{\bb} \nabla_{{\bf x}_1} P_{\bb \rr} \right) \mathrm{d}{\bf x}_2\\
& + \int_{\Omega_{\epsilon}({\bf x}_1)}  \nabla_{{\bf x}_2} \cdot  \left( D_{\bb}  \nabla_{{\bf x}_2} P_{\bb \bb} + D_{\rr} \nabla_{{\bf x}_2} P_{\bb \rr} \right)  \mathrm{d}{\bf x}_2 \\
& + k_\rr r - k_\bb b.
\end{aligned}
\end{align}
By using the divergence and the Reynolds transport theorems, we simplify equation (\ref{SS_blue_CI_1}) \review{to give the following integro-partial differential equation}
\begin{subequations} \label{int_br}
\begin{align} \label{SS_pde_CI_blue}
\begin{aligned}
\partial_t b & =   D_{\bb} \nabla_{{\bf x}_1}^2 b  + k_\rr r - k_\bb b \\
& \ + \! \int_{\partial B_{\epsilon}({\bf x}_1)} \! -\big[2D_{\bb} \nabla_{{\bf x}_1} (P_{\bb \bb}+P_{\bb \rr} )  \\
& \qquad \qquad \qquad + (D_{\bb}-D_{\rr}) \nabla_{{\bf x}_2} P_{\bb \rr} \big] \cdot \mathbf{n_2} \mathrm{d}S_{{\bf x}_2} ,   
\end{aligned}
\end{align}
where $\partial B_{\epsilon}({\bf x}_1)$ is the collision surface for a sphere of radius $\epsilon$ at position ${\bf x}_1$. Similarly we can derive the following equation for $r({\bf x}_1,t)$:
\begin{align}  \label{SS_pde_CI_red}
\begin{aligned}
\partial_t r & =   D_{\rr} \nabla_{{\bf x}_1}^2 r  - k_\rr r + k_\bb b\\
& \ + \! \int_{\partial B_{\epsilon}({\bf x}_1)}  \! - \big[ 2D_{\rr} \nabla_{{\bf x}_1} (P_{\rr \bb}+P_{\rr \rr} )  \\
&\qquad \qquad \qquad + (D_{\rr}-D_{\bb}) \nabla_{{\bf x}_2} P_{\rr \bb} \big] \cdot \mathbf{n_2} \mathrm{d}S_{{\bf x}_2}.
\end{aligned}
\end{align} 
\review{Equations \eqref{int_br} are complemented with zero-flux boundary conditions on $\partial \Omega$ and initial conditions  $b({\bf x}_1,0)=b_0({\bf x}_1)$ and $r({\bf x}_1,0)=r_0({\bf x}_1)$.} The integrals \eqref{int_br} are on the contact surface of two interacting particles, where the particle positions are correlated. As a result, a closure approximation of the type $P_{\bb \bb}({\bf x}_1,{\bf x}_2,t) = b({\bf x}_1,t)b({\bf x}_2,t)$ is not suitable.  Instead we determine the collision integrals systematically, using the method of matched asymptotic expansions.
\end{subequations}

\subsection{Matched asymptotic expansions for the densities \texorpdfstring{$P_{\bb \bb}$}{Pbb} and \texorpdfstring{$P_{\rr \rr}$}{Prr}} \label{Coll_C}

We define two regions in configuration space, the inner region where the two particles are close to each other, $\|{\bf x}_1-{\bf x}_2\| \sim \epsilon$, and the outer region where they are far apart, $\|{\bf x}_1-{\bf x}_2\| \gg \epsilon$. To evaluate the collision integral in equation (\ref{SS_pde_CI_blue}) we must first calculate an asymptotic expansion for $P_{\bb \bb}({\bf x}_1,{\bf x}_2,t)$ in the inner region. In the outer region the positions of the two particles are assumed to be uncorrelated at leading order. Therefore
\begin{equation} \label{out_bb}
P_{\bb \bb}^{\text{out}} =  q_{\bb}({\bf x}_1,t) q_{\bb}({\bf x}_2,t) + \epsilon P_{\bb \bb}^{\text{out},(1)}({\bf x}_1,{\bf x}_2,t) + \cdots,
\end{equation}
for some function $q_{\bb}$. In the inner region, we introduce the inner variables ${\bf x}_1 = \hat{{\bf x}}_1$ and ${\bf x}_2 = \hat{{\bf x}}_1 + \epsilon \hat{{\bf x}}$ and define $\hat{P}_{\bb \bb}(\hat{{\bf x}}_1,\hat{{\bf x}},t) = P_{\bb \bb}({\bf x}_1,{\bf x}_2,t)$. Then equation (\ref{SS_PPDE_bb}) becomes
\begin{subequations} \label{inner_problem}
\begin{align}
\epsilon^2 \partial_t \hat{P}_{\bb \bb}  = \ & 2D_{\bb} \nabla_{\hat{{\bf x}}}^2 \hat{P}_{\bb \bb} - 2 \epsilon D_{\bb}  \nabla_{\hat{{\bf x}}_1} \cdot \nabla_{\hat{{\bf x}}} \hat{P}_{\bb \bb}  \label{SS_Pbb_0} \\
& + \epsilon^2 \big[ D_{\bb} \nabla_{\hat{{\bf x}}_1}^2 \hat{P}_{\bb \bb} + k_\rr ( \hat{P}_{\bb \rr} + \hat{P}_{\rr \bb} ) - 2k_\bb \hat{P}_{\bb \bb} \big]. \nonumber
\end{align}
The zero-flux boundary condition (\ref{SS_PPDE_BC}) becomes
\begin{equation}\label{SS_BC_bb}
2\hat{{\bf x}} \cdot \nabla_{\hat{{\bf x}}} \hat{P}_{\bb \bb} = \epsilon  \hat{{\bf x}} \cdot \nabla_{\hat{{\bf x}}_1} \hat{P}_{\bb \bb}, \qquad \text{on} \qquad \|\hat{{\bf x}}\|=1.  
\end{equation}
To match the inner and outer regions we expand the outer solution \eqref{out_bb} in the inner variables to give
\begin{align}
\begin{aligned}
\hat P_{\bb \bb} \sim \ & q_{\bb}^2(\hat{{\bf x}}_1,t) + \epsilon \big[ q_{\bb}(\hat{{\bf x}}_1,t) \hat{{\bf x}} \cdot \nabla_{\hat{{\bf x}}_1} q_{\bb}(\hat{{\bf x}}_1,t) \\
& + P_{\bb \bb}^{\text{out},(1)}(\hat{{\bf x}}_1,\hat{{\bf x}}_1,t) \big],
\end{aligned}
\end{align}
as $\|\hat{{\bf x}}\| \to \infty$. 
\end{subequations}
We seek a regular power series expansion of the form $\hat{P}_{\bb \bb} = \hat{P}_{\bb \bb}^{(0)} + \epsilon \hat{P}_{\bb \bb}^{(1)} + \cdots$. Substituting in \eqref{inner_problem} and equating terms of zeroth order we obtain the following problem
\begin{alignat*}{2}
\nabla_{\hat{{\bf x}}}^2 \hat{P}_{\bb \bb}^{(0)} &= 0, & &   \\
\hat{{\bf x}} \cdot \nabla_{\hat{{\bf x}}} \hat{P}_{\bb \bb}^{(0)} &= 0, &\qquad &\text{on} \qquad \|\hat{{\bf x}}\|=1, \\
\hat{P}_{\bb \bb}^{(0)}& \sim q_{\bb}^2(\hat{{\bf x}}_1,t), &\qquad &\text{as} \qquad \|\hat{{\bf x}}\| \rightarrow \infty, 
\end{alignat*}
with solution $\hat{P}_{\bb \bb}^{(0)} = q_{b}^2(\hat{{\bf x}}_1,t)$. Similarly, the first order problem for $\hat{P}_{\bb \bb}^{(1)}$ is
\begin{alignat*}{2}
\nabla_{\hat{{\bf x}}}^2 \hat{P}_{\bb \bb}^{(1)} &=  0,  \\
\hat{{\bf x}} \cdot \nabla_{\hat{{\bf x}}} \hat{P}_{\bb \bb}^{(1)} & =  q_{\bb}(\hat{{\bf x}}_1,t) \hat{{\bf x}} \cdot \nabla_{\hat{{\bf x}}_1} q_{\bb}(\hat{{\bf x}}_1,t), \quad \text{on}  \ \|\hat{{\bf x}}\|=1, \\
\hat{P}_{\bb \bb}^{(1)} & \sim q_{\bb}(\hat{{\bf x}}_1,t) \hat{{\bf x}} \cdot \nabla_{\hat{{\bf x}}_1} q_{\bb}(\hat{{\bf x}}_1,t) \nonumber \\
& \quad + P_{\bb \bb}^{\text{out},(1)}(\hat{{\bf x}}_1,\hat{{\bf x}}_1,t), \quad \text{as} \ \|\hat{{\bf x}}\| \rightarrow \infty, 
\end{alignat*}
with solution $\hat{P}_{\bb \bb}^{(1)} = q_{\bb}(\hat{{\bf x}}_1,t) \hat{{\bf x}} \cdot \nabla_{\hat{{\bf x}}_1} q_{\bb}(\hat{{\bf x}}_1,t) + P_{\bb \bb}^{\text{out},(1)}(\hat{{\bf x}}_1,\hat{{\bf x}}_1,t)$. We conclude that, to first order in $\epsilon$, the expansion of $P_{\bb \bb}({\bf x}_1,{\bf x}_2,t)$ in the inner region is
\begin{align}
\begin{aligned}
\hat{P}_{\bb \bb} \sim \ & q_{\bb}^2(\hat{{\bf x}}_1,t) + \epsilon \big[ q_{\bb}(\hat{{\bf x}}_1,t) \hat{{\bf x}} \cdot \nabla_{\hat{{\bf x}}_1} q_{\bb}(\hat{{\bf x}}_1,t) \\
& + P_{\bb \bb}^{\text{out},(1)}(\hat{{\bf x}}_1,\hat{{\bf x}}_1,t) \big]. 
\end{aligned}
\label{Pbb}
\end{align}
Similarly, to first order we can write the expansion of $P_{\rr \rr}({\bf x}_1,{\bf x}_2,t)$ in the inner region as
\begin{align}
\begin{aligned}
\hat{P}_{\rr \rr} \sim q_{\rr}^2(\hat{{\bf x}}_1,t) + \epsilon \big[ & q_{\rr}(\hat{{\bf x}}_1,t) \hat{{\bf x}} \cdot \nabla_{\hat{{\bf x}}_1} q_{\rr}(\hat{{\bf x}}_1,t) \\
& + P_{\rr \rr}^{\text{out},(1)}(\hat{{\bf x}}_1,\hat{{\bf x}}_1,t) \big].
\end{aligned}
\end{align}

\subsection{Matched asymptotic expansions for the densities \texorpdfstring{$P_{\bb \rr}$}{Pbr} and \texorpdfstring{$P_{\rr \bb}$}{Prb}}

In order to evaluate the collision integral in (\ref{SS_pde_CI_blue}) we require an asymptotic expansion for $P_{\bb \rr}({\bf x}_1,{\bf x}_2,t)$ in the inner region. In the outer region, our assumption of independence enables us to write
\begin{equation}\label{SS_out_br}
P_{\bb \rr}^{\text{out}} = q_{\bb}({\bf x}_1,t)q_{\rr}({\bf x}_2,t) + \epsilon P_{\bb \rr}^{\text{out},(1)}({\bf x}_1,{\bf x}_2,t) + \cdots,
\end{equation}
for some functions $q_{\bb}$ and $q_{\rr}$. 

\begin{subequations}\label{innerPbr}
In the inner region, we define $ \hat{P}_{\bb \rr}(\hat{{\bf x}}_1,\hat{{\bf x}},t) = P_{\bb \rr}({\bf x}_1,{\bf x}_2,t)$ and transform equation (\ref{SS_PPDE_br}) into inner variables as before, yielding
\begin{align}
\epsilon^2 \partial_t \hat{P}_{\bb \rr} &=(D_{\bb}+D_{\rr})\nabla_{\hat{{\bf x}}}^2 \hat{P}_{\bb \rr} - 2\epsilon D_{\bb} \nabla_{\hat{{\bf x}}_1} \cdot \nabla_{\hat{{\bf x}}} \hat{P}_{\bb \rr}  \label{SS_PPDE_innek_br}
\\
& + \epsilon^2 \big[ D_{\bb} \nabla_{\hat{{\bf x}}_1}^2 \hat{P}_{\bb \rr} + k_\bb \hat{P}_{\bb \bb} + k_\rr \hat{P}_{\rr \rr} - (k_\bb \!+ \! k_\rr)\hat{P}_{\bb \rr} \big]. \nonumber
\end{align}
The  boundary condition (\ref{SS_PPDE_BC}) becomes 
\begin{equation}
\hat{{\bf x}} \cdot \nabla_{\hat{{\bf x}}} \hat{P}_{\bb \rr} = \epsilon \dfrac{D_{\bb}}{D_{\bb} + D_{\rr} } \hat{{\bf x}} \cdot \nabla_{\hat{{\bf x}}_1} \hat{P}_{\bb \rr},  \label{SS_BC_br}
\end{equation}
on $\|\hat{{\bf x}}\|=1$. 
Expanding (\ref{SS_out_br}) in inner variables gives the matching condition
\begin{align} \label{matching_switch}
\hat P_{\bb \rr} \sim \ & q_{\bb}(\hat{{\bf x}}_1,t) q_{\rr}(\hat{{\bf x}}_1,t) \\&+ \epsilon [  q_{\bb}(\hat{{\bf x}}_1,t) \hat{{\bf x}} \cdot \nabla_{\hat{{\bf x}}_1} q_{\rr}(\hat{{\bf x}}_1,t)  + P_{\bb \rr}^{\text{out},(1)}(\hat{{\bf x}}_1,\hat{{\bf x}}_1,t)], \nonumber
\end{align}
as $\|\hat{{\bf x}}\|\to \infty$. 
\end{subequations}

As before, we seek a solution of the form $\hat{P}_{\bb \rr} = \hat{P}_{\bb \rr}^{(0)} + \epsilon \hat{P}_{\bb \rr}^{(1)} + \cdots$. Repeating the procedure outlined in the previous section, we find that the inner region solution for a blue and a red particle is
\begin{align}
\begin{aligned}
\hat{P}_{\bb \rr} \sim \ &  q_{\bb}q_{\rr} + \epsilon q_{\bb} \hat{{\bf x}} \cdot \nabla_{\hat{{\bf x}}_1} q_{\rr} + \epsilon P_{\bb \rr}^{\text{out},(1)}(\hat{{\bf x}}_1,\hat{{\bf x}}_1,t) \\& + \dfrac{\epsilon \hat{{\bf x}} \cdot ( D_{\rr} q_{\bb}\nabla_{\hat{{\bf x}}_1} q_{\rr} - D_{\bb} q_{\rr} \nabla_{\hat{{\bf x}}_1} q_{\bb} )}{(D_{\bb}+D_{\rr})(d-1)\|\hat{{\bf x}}\|^d}  ,
\end{aligned} \label{Pbr}
\end{align}
\review{where $q_{\bb}$ and $q_{\rr}$ are evaluated at $(\hat{{\bf x}}_1,t)$.}
Similarly, to first order we find that the expansion of $P_{\rr \bb}({\bf x}_1,{\bf x}_2,t)$ in the inner region is
\begin{align}
\begin{aligned}
\hat{P}_{\rr \bb} \sim \ &  q_{\rr}q_{\bb} + \epsilon q_{\rr} \hat{{\bf x}} \cdot \nabla_{\hat{{\bf x}}_1} q_{\bb}
 + \epsilon P_{\rr \bb}^{\text{out},(1)}(\hat{{\bf x}}_1,\hat{{\bf x}}_1,t) \\ 
 & + \dfrac{\epsilon \hat{{\bf x}}  \cdot (D_{\bb} q_{\rr}\nabla_{\hat{{\bf x}}_1} q_{\bb} - D_{\rr} q_{\bb} \nabla_{\hat{{\bf x}}_1} q_{\rr})}{(D_{\bb}+D_{\rr})(d-1)\|\hat{{\bf x}}\|^d}.
\end{aligned}
\end{align}

\subsection{Evaluating the collision integral}

\review{Since the collision integral, which we shall denote by  $I$, in \eqref{SS_pde_CI_blue} is defined over the boundary surface $\partial B_{\epsilon}({\bf x}_1)$, we use the inner solutions $\hat P_{\bb \bb}$ and $\hat P_{\bb \rr}$ in order to evaluate it. We first transform $I$ into inner variables and 
use the boundary conditions (\ref{SS_BC_bb}) and (\ref{SS_BC_br}) yielding}
\begin{equation}\label{SS_CI}
I = \epsilon^{d-1} D_{\bb} \int_{\partial B_1({\bf 0})} \nabla_{\hat{{\bf x}}_1} (\hat{P}_{\bb \bb}+\hat{P}_{\bb \rr}) \cdot \hat{{\bf x}} \, \mathrm{d} S_{\hat{{\bf x}}}.
\end{equation}
Inserting  the inner region solutions (\ref{Pbb}) and (\ref{Pbr}) into (\ref{SS_CI}), we find 
\begin{align} \label{integ_com}
\begin{aligned}
I \sim  \epsilon^d D_{\bb}\nabla_{\bfx_1} \cdot ( \omega q_{\bb}\nabla_{\bfx_1} q_{\bb}  +  \xi_{\bb}  q_{\bb} \nabla_{{{\bf x}}_1} q_{\rr} - \eta_{\bb} q_{\rr} \nabla_{{{\bf x}}_1} q_{\bb} ),  
\end{aligned}
\end{align}
where the constants $\omega$, $\xi_{\bb}$ and $\eta_{\bb}$ are given in (\ref{parameters}). Using the  normalization condition $\int_{\Omega_{\epsilon}^2} \left( P_{\bb \bb} + P_{\bb \rr} + P_{\rr \bb} + P_{\rr \rr} \right) \ud \bfx_1 \bfx_2= 1$,  we find that $q_{\bb}({\bf x}_1,t) = b({\bf x}_1,t) + O(\epsilon^2)$ and $q_{\rr}({\bf x}_1,t) = r({\bf x}_1,t) + O(\epsilon^2)$.
This allows us to replace $q_{\bb}$ and $q_{\rr}$ in \eqref{integ_com} by  $b$ and $r$, respectively.

Substituting the asymptotic expansion for the collision integral \eqref{integ_com} into equation (\ref{SS_pde_CI_blue}) we arrive at the following  equation for the blue particles' density $b(\bfx,t)$ (\review{after replacing $\bfx_1$ by $\bfx$}):
\begin{align*}
\begin{aligned}
\partial_t b = \ & \nabla_{{\bf x}} \cdot \big [D_{\bb} ( 1 + \omega \epsilon^d b ) \nabla_{{\bf x}} b  + \epsilon^d D_{\bb} ( \xi_{\bb} b \nabla_{{\bf x}} r - \eta_{\bb} r \nabla_{{\bf x}} b ) \big] \\
& + k_\rr r - k_\bb b.
\end{aligned}
\end{align*}
A similar calculation on (\ref{SS_pde_CI_red}) yields an equation for the density of red particles  $r(\bfx,t)$. 

For the case of $N$ particles there is a contribution from every potential pairwise interaction. This allows us to write down the population-level equation for the general case, where we reintroduce the drift contributions, as follows
\begin{subequations} \label{model_spo}
\begin{align}
\label{crossreactiondiffusion}
\begin{aligned}
\partial_t \begin{pmatrix} b \\ r \end{pmatrix}= \ &  \nabla_{{\bf x}} \cdot \left[ {\mathcal D} \, \nabla_{{\bf x}} \! \begin{pmatrix} b \\ r \end{pmatrix} - {\mathcal F} \begin{pmatrix} b \\ r \end{pmatrix}
 \right] + \begin{pmatrix} k_\rr r - k_\bb b\\ k_\bb b - k_\rr r \end{pmatrix} , 
\end{aligned}
\end{align}
\review{with zero-flux boundary conditions on $\partial \Omega$, and initial conditions $b({\bf x},0)=b_0({\bf x})$ and $r({\bf x},0)=r_0({\bf x})$.} \review{The diffusion matrix is}
\begin{align}\label{diff_matrix_switching}
 {\mathcal D} = \text{diag}\begin{pmatrix} D_{\bb}\\ D_{\rr} \end{pmatrix} + (N-1) \epsilon^d \begin{pmatrix} D_{\bb}(  \omega b   -  \eta_{\bb} r ) , \quad D_{\bb}  \xi_{\bb} b   \\ D_{\rr} \xi_{\rr} r, \quad D_{\rr}( \omega r   -   \eta_{\rr} b ) \end{pmatrix}, 
\end{align}
and the drift matrix is
 \begin{align}\label{drift_matrix_switching}
 {\mathcal F} = \text{diag} \begin{pmatrix} {\bf f}_\bb \\ {\bf f}_\rr \end{pmatrix} + (N-1) \epsilon^d \begin{pmatrix} {\bf 0}, \quad  \eta_{\bb} ( {\bf f}_\rr - {\bf f}_\bb ) b \\ \eta_{\rr} ( {\bf f}_\bb - {\bf f}_\rr ) r , \quad {\bf 0} \end{pmatrix}.
 \end{align}
The diffusion and drift matrices (\ref{diff_matrix_switching}) and (\ref{drift_matrix_switching}) can be compared to the matrices (\ref{diff_matrix_no_switching}) and (\ref{drift_matrix_no_switching}) corresponding to the case of two subpopulations without switching \cite{bruna2012multiple}. \review{In that case,} the numbers of blue and red particles, $N_{\bb}$ and $N_{\rr}$, were fixed and appeared explicitly in the matrices. As now there is a non zero probability of being either blue or red we see contributions from all $\left( N-1 \right)$ possible collisions in (\ref{diff_matrix_switching}) and (\ref{drift_matrix_switching}). 
\end{subequations}

\subsection{Numerical results}

\label{CollNR}

In this section we present numerical simulations of the discrete stochastic model \eqref{sde_switch} and the population-based model \eqref{model_spo}. The discrete model was simulated  using the software package Smoldyn \cite{andrews2004stochastic, smoldyn} via the standard Euler--Maruyama method \cite{erban2007BCs}.

If two particles are found to be overlapping then we incorporate volume exclusion by using Smoldyn's ballistic collision update rule. This method assigns to each particle a post-collision velocity by comparing both particles' current and previous positions. Then it uses conservation of momentum to compute their positions had they undergone an elastic collision. 

When considering spontaneous switching of particles between subpopulations, at each timestep in the algorithm we generate a random number $\rho$ uniformly on the interval $[0,1]$. A blue particle becomes red if $\rho < k_\bb \Delta t$. Similarly, a red particle becomes blue if $\rho < k_\rr \Delta t$. In order for this implementation to be accurate we assume $\Delta t$ to be small enough such that $k_\bb \Delta t \ll 1$ and  $k_\rr \Delta t \ll 1$, and so are good approximations to the probability of a switch occurring in a time interval of length $\Delta t$. We also require that the timestep is small enough that each particle moves, on average, a distance much less than the particle's diameter. This requirement ensures that most collisions are detected. 

We solved the PDEs numerically using the method of lines. In this technique only the spatial variables are discretized and time is viewed as a continuous variable, resulting in a system of Ordinary Differential Equations (ODEs) that are solved using MATLAB's inbuilt \texttt{ode15s} solver (which performs the discretization of the time variable adaptively). 

\begin{figure}
\begin{center}
	\includegraphics[width=0.49\textwidth]{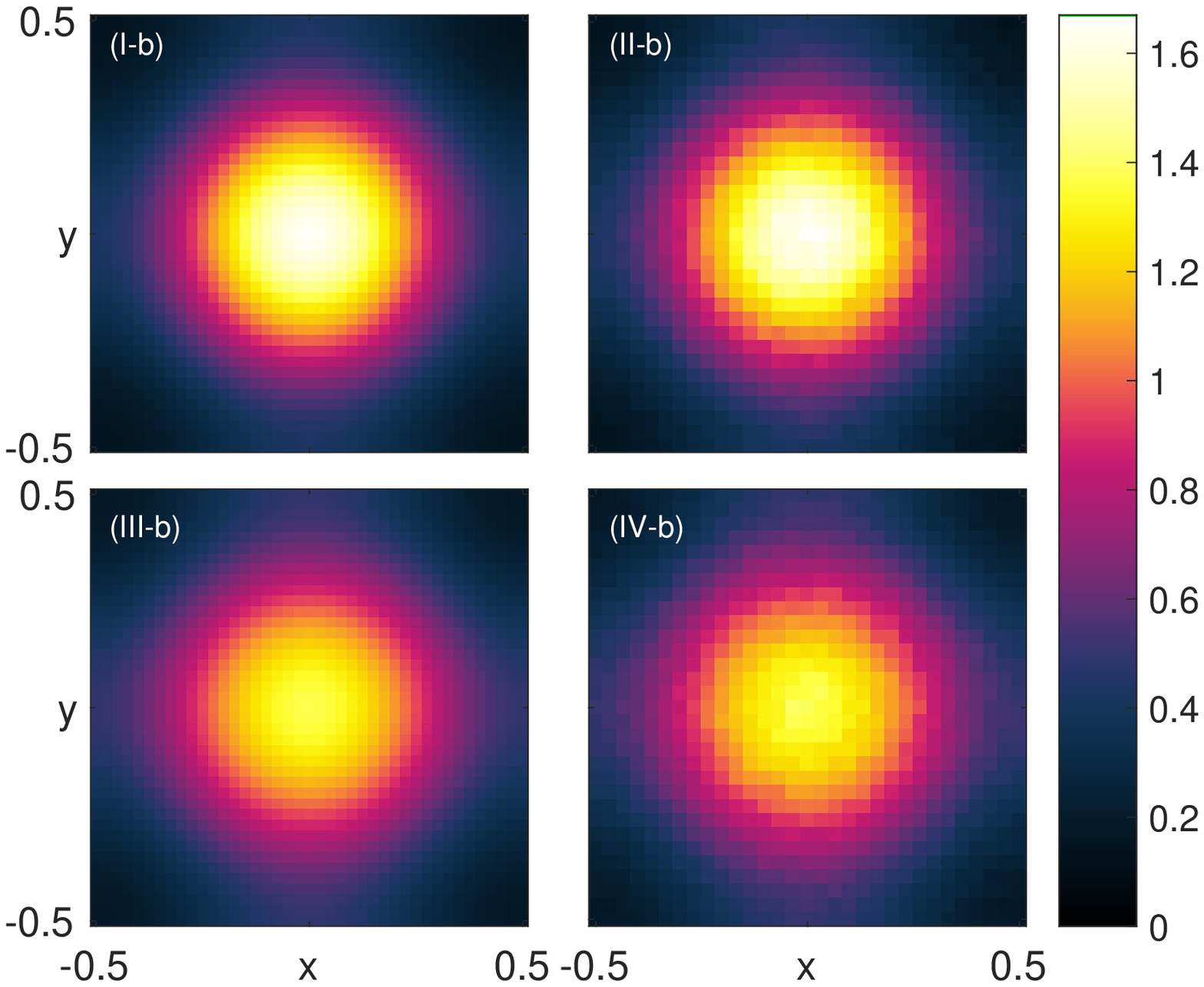}  \includegraphics[width=0.49\textwidth]{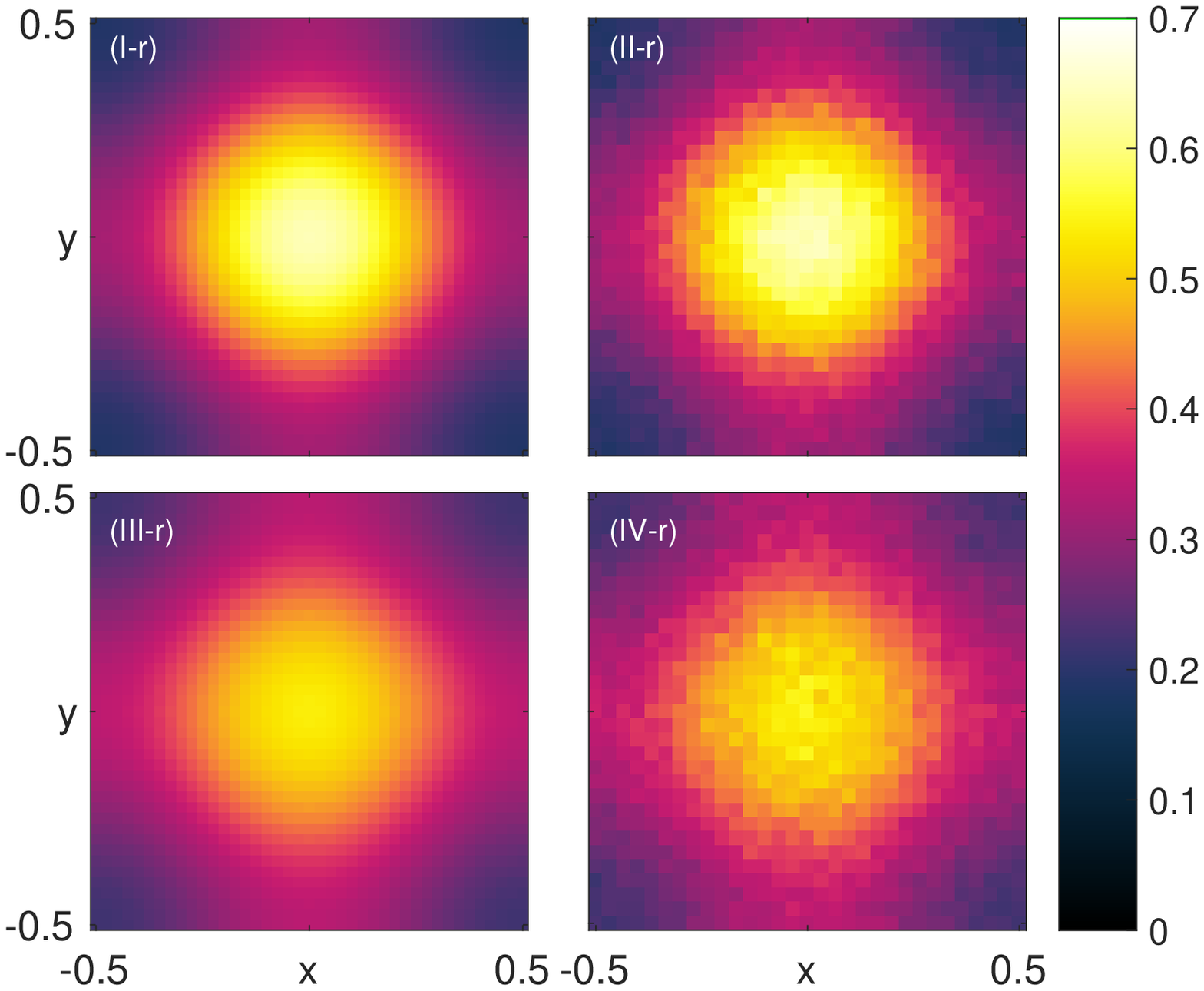}
\end{center}
\caption{ Population-level equations (\ref{crossreactiondiffusion}) for both blue ([I-IV]-b) and red ([I-IV]-r) particles at time $t=0.05$, with data being initially Gaussian with zero mean and standard deviation $0.09$. (I) Solutions to equations (\ref{crossreactiondiffusion}) for point particles, $\epsilon = 0$. (II) Histograms for point particles, $\epsilon = 0$. (III) Solutions to equations (\ref{crossreactiondiffusion}) for finite size particles, $\epsilon = 0.01$. (IV) Histograms for finite size particles, $\epsilon = 0.01$. The data was calculated from $10^4$ stochastic simulations with $\Delta t = 10^{-5}$. Parameter values: $D_{\bb}=0.5$, $D_{\rr}=1$, $k_\bb=k_\rr=10$ and $N=400$. Initial conditions were $N_{\bb}(0)=350$ and $N_{\rr}(0)=50$.}
\label{fig:fig1}
\end{figure}

Initially we have $350$ blue particles and $50$ red particles in the domain $\Omega = [-1/2,1/2]^2$. We select an initial Gaussian density of particles with zero mean and a standard deviation of $0.09$ for both the blue and red subpopulations. We set $D_{\bb}=0.5$ and $D_{\rr}=1$ for the diffusion coefficients of the blue and red populations respectively, as well as switching parameters $k_\bb = k_\rr =10$. We ignore drift terms, setting ${\bf f}_\bb = {\bf f}_\rr = {\bf 0}$. The data from the stochastic simulations of the SDEs (\ref{SDE_SSb}) and (\ref{SDE_SSr}) were generated using $10^4$ realizations, which resulted in $4 \times 10^6$ individual trajectories, where a timestep of $\Delta t = 10^{-5}$ was used. \review{The stochastic simulations used to generate the results presented in this section were performed on an \texttt{AMD FX(tm)-4350} CPU, with an advertised processor speed of $4.2$ GHz. The total CPU time needed to generate the trajectories seen in Figure \ref{fig:fig1} is approximately $120$ hours, whereas the PDEs  \eqref{model_spo} are solved numerically within $13$ seconds. }

\begin{figure}
\begin{center}
	\includegraphics[width=.45\textwidth]{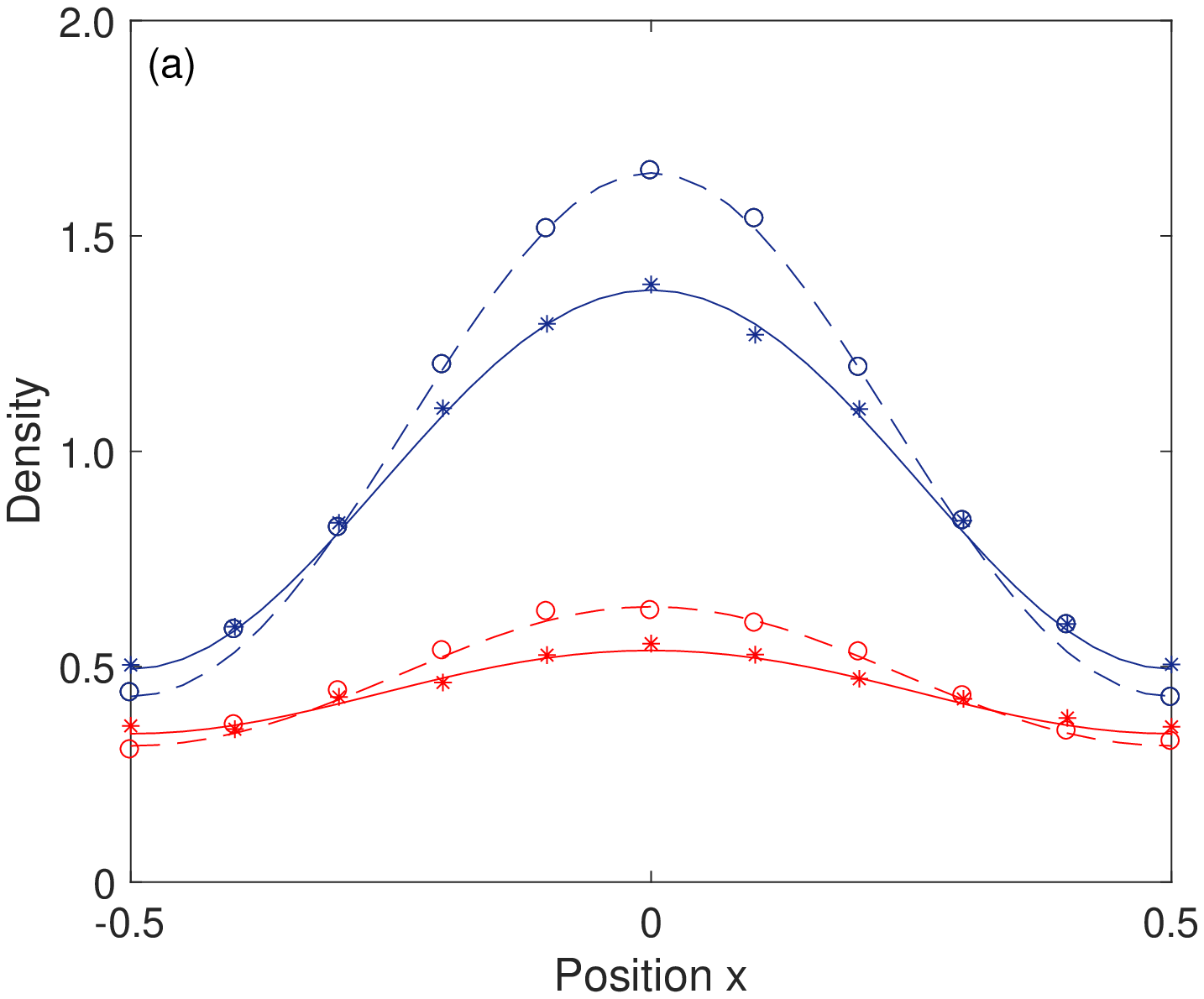} \quad \includegraphics[width=.45\textwidth]{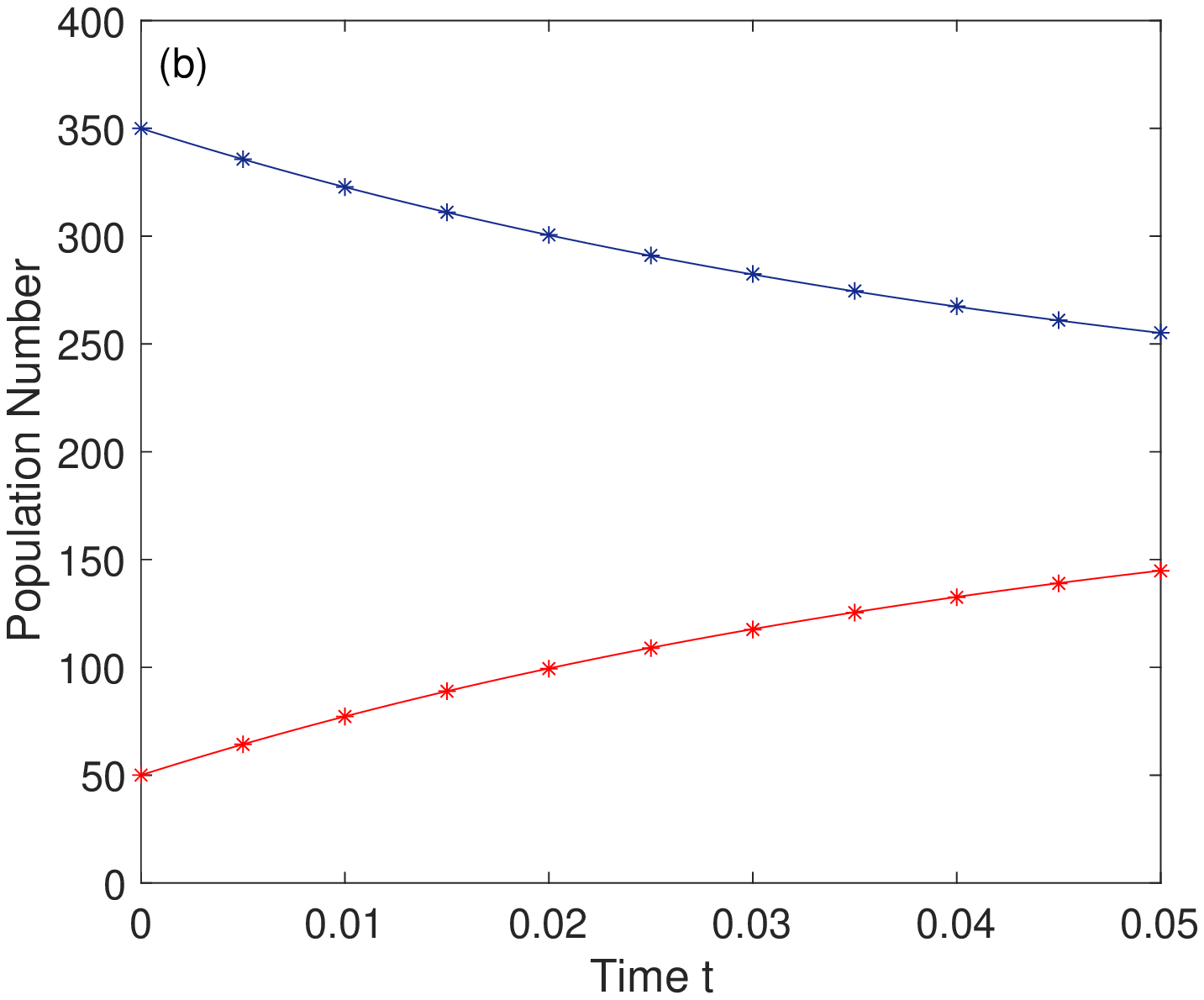}
\end{center}
\caption{ (a) A slice of the histograms in Figure \ref{fig:fig1} at $y=0$. (b) The evolution of the blue and red population for $0 \leq t \leq 0.05$ for finite size particles ($\epsilon=0.01$). Solid (dashed) lines represent the PDE and ODE solutions for finite size (point) particles. Asterisks (circles) represent data from stochastic simulations for finite size (point) particles. Parameter values \review{as in Figure \ref{fig:fig1}}.}
\label{fig:fig2}
\end{figure}

Figure \ref{fig:fig1} shows the densities $b(\bfx,t)$ and $r(\bfx,t)$ of the blue and red particles respectively at time $t=0.05$. Panels (I) and (II) respectively show the PDE solution and the histogram produced from the simulated trajectories for point particles ($\epsilon = 0$), whereas panels (III) and (IV) respectively show the same for finite size particles of diameter $\epsilon = 0.01$. Figure \ref{fig:fig2}(a) shows a one dimensional slice of the histograms in Figure \ref{fig:fig1} for a clearer comparison of the PDEs and the stochastic data. We observe good agreement between the particle-based model and the derived population-level equations. \review{The relative $L_2$-norm errors in the densities (between the solution of the PDE model and the histograms of the stochastic particle-based model) are, from top to bottom in Figure \ref{fig:fig2}(a), $2.5\%, 1.7\%, 2.6\%$ and $2.7\%$.}

For further quantitative comparison we consider how the population numbers, $N_{\bb}(t)$ and $N_{\rr}(t)$, evolve over time. Recall that $\int_{\Omega} b({\bf x},t) \mathrm{d}{\bf x} = N_{\bb}(t)/N$ and $\int_{\Omega} r({\bf x},t) \mathrm{d}{\bf x} = N_{\rr}(t)/N$. We can then integrate equations (\ref{crossreactiondiffusion}) over the domain $\Omega$ to obtain, using zero-flux boundary conditions, the following ODE
\begin{equation} \label{ODE}
N_{\bb}'(t) = k_\rr N_{\rr}(t) - k_\bb N_{\bb}(t), \quad N_{\rr}(t)=N-N_{\bb}(t).
\end{equation}
Figure \ref{fig:fig2}(b) shows the solution of (\ref{ODE}) and compares them with results from the stochastic simulations. We only plot results for finite size particles but note that the curves for point particles are identical. That is, the finite size particles do not effect the evolution of the population number when the switching rates are constant on $\Omega$, and so the ODEs are exactly the same as if we had considered the non-spatial model originally. 

\subsection{\review{Application to a hybrid model of chemotaxis}} 
\label{chemo}
\review{We now consider a three species model of chemotaxis as an application of our proposed framework. As before we consider two populations of finite-size particles, blue and red, which we shall refer to as cells. We introduce a third species which shall act as a chemoattractant that is produced by the red cells and acts as a signaling chemical towards which the blue cells migrate. The size of chemoattractant molecules is negligible in comparison to the size of the cells which produce them. As such we consider the chemoattractant as a continuum field described by a reaction-diffusion PDE. This yields a hybrid modeling framework, whereby some species are described as a continuum and some species as a set of particles \cite{Franz:2013ds}. Hybrid chemotaxis models have been developed in the literature \cite{Dallon:1997be, McLennan:2012jb}. In \cite{Dallon:1997be}, a population of identical cells that only interact via the chemical signal (that is, cells are represented as point particles) is considered. The model in \cite{McLennan:2012jb} describes cell migration during embryonic development and includes two types of cells (``leaders'' and ``followers''). Cells undergo a position-jump process (with jumps of a fixed distance) and volume exclusion between cells is included as a rejection mechanism (cells are modeled as hard discs and a jump is aborted if a cell attempts to move to a position where it overlaps with another cell). }

\review{In this section, we use our model for two reacting species of hard-sphere particles in the context of hybrid modeling of chemotaxis. In contrast to \cite{McLennan:2012jb}, our model is for Brownian particles and includes a more accurate description of excluded-volume interactions (not biasing the motion of cells by aborting moves that would lead to overlaps \cite{Strating:1999uw}).  The red cells secrete chemoattractant at a constant rate $\gamma$, while the blue cells undergo biased motion up spatial gradients of the chemoattractant. The chemoattractant diffuses throughout the domain without interacting with the cells and degrades at a constant rate $\mu$. An additional mechanism for the loss of chemoattractant due to consumption by blue cells could be considered here. However, under the assumption that the dominant mechanism for the loss of chemoattractant is natural decay \cite{Grima::PRL05}, the additional consumption term is neglected. We assume cells move in a two-dimensional domain $\Omega$ and that the chemical is allowed anywhere in the domain (to represent cells in contact with the substrate and the chemical diffusing in the space above). We denote the position and the color of the $i$th cell at time $t$ as ${\bf X}_i(t)$ and $S_i(t) \in\{\bb,\rr\}$ respectively. We denote by $C({\bf x},t | \vec{x},\vec{s} )$ the concentration of the chemoattractant at position ${\bf x} \in \Omega$ at time $t$ given a configuration state $\vec{x}=\{{\bf X}_1(t),\ldots,{\bf X}_N(t)\}$ and $\vec{s}=\{S_1(t),\ldots,S_N(t) \}$. The microscopic model for the blue cells ($S_i(t) = \bb$) is}
\begin{subequations} \label{sde_chemo}
\begin{align}
\eqreview{\mathrm{d}} {\bf X}_i(t) &\eqreview{ = \sqrt{2D_{\bb}}} \mathrm{d} {\bf W}_i(t) + \chi \nabla C({\bf X}_i(t),t) \mathrm{d}t\label{SDE_Chemob}, 
\end{align}
\review{where $\chi \geq 0$ is the affinity of the blue cells for the chemoattractant. Red cells ($S_i(t) = \rr$) follow a simple Brownian motion as before}
\begin{align}
\eqreview{\mathrm{d}} {\bf X}_i(t) &\eqreview{ = \sqrt{2D_{\rr}}} \mathrm{d}{\bf W}_i(t),  \label{SDE_Chemor}
\end{align}
\review{for $1 \leq i \leq N$. The evolution of the chemoattractant $C({\bf x},t|\vec{x},\vec{s})$ is governed by the following PDE}
\begin{align}
\label{pde_c}
\eqreview{\partial_t C= } \eqreview{D_c \nabla^2 C + \gamma \! \sum_{i\in \mathcal I_{\rr}(t)} K_h({\bf x}-{\bf X}_i(t)) } - \mu C,
\end{align}
\review{for $\bfx \in \Omega$, with zero-flux boundary conditions on $\partial \Omega$, initial conditions $C({\bf x},0)=C_0({\bf x})$. Here $D_c$ is the diffusion coefficient for the chemoattractant and $K_h(\bfx)$ denotes a Gaussian kernel of bandwidth $h$, $K_h(\bfx) = \exp(-\| \bfx\|^2/(2h^2))/(2 \pi h^d)$. We set the bandwidth of the Kernel Density estimation equal to $\epsilon/2$ to model the fact that cells will produce chemoattractant over a region related to their radius, rather than at a single point ${\bf X}_i$, as a term  of the form $\delta({\bf x}-{\bf X}_i(t))$ instead of $K_h$ in \eqref{pde_c} would represent. We note that, even in models with point cells, KDE is used to approximate the sum of delta functions in hybrid models \cite{Franz:2013ds}.}

\review{Finally, cells change color according to the reactions}
\begin{equation}
\review{B \xrightarrow{k_\bb C} R, \qquad R \xrightarrow{k_\rr} B}. \label{Chemo_reactions}
\end{equation}
\review{Blue cells switch color at a rate proportional to the concentration of chemoattractant at their current position $X_i(t)$, and red cells switch color at a constant rate $k_\rr$ as before. Similar to Section \ref{Particle-based}, we can formulate the microscopic model in terms of a high-dimensional Fokker--Planck equation and introduce the population-level equations $b({\bf x},t)$ and $r({\bf x},t)$ as in \eqref{br_densities}. Additionally introducing the population-level equation for the chemoattractant $c({\bf x},t) =  \int_{\Omega_{\epsilon}^N} \sum_{\vec{s} \in \mathcal{C}^N} C({\bf x},t | \vec{x}, \vec{s}) \mathrm{d}\vec{x}$ allows us to write down the following population-level reaction-diffusion equations}
\end{subequations}
\begin{subequations} \label{model_chemo}
\begin{align}
\begin{aligned}
\eqreview{\partial_t \begin{pmatrix} b \\ r \\ c \end{pmatrix} }=\  & \eqreview{\nabla_{{\bf x}} }\cdot \left[ \widehat {\mathcal D} \, \nabla_{{\bf x}} \! \begin{pmatrix} b \\ r \\ c \end{pmatrix} - \widehat {\mathcal F} \begin{pmatrix} b \\ r \\ c \end{pmatrix}
 \right] \\
 &\eqreview{+} \begin{pmatrix} k_\rr r - k_\bb c b\\ k_\bb c b - k_\rr r \\ \gamma N r  - \mu c \end{pmatrix}, 
\end{aligned}
\end{align}
\review{with zero-flux boundary conditions on $\partial \Omega$ and initial conditions $b({\bf x},0)=b_0({\bf x})$, $r({\bf x},0)=r_0({\bf x})$ and $c({\bf x},0)=c_0({\bf x})$. The diffusion matrix is}
\begin{align}\label{diff_matrix_chemo}
\eqreview{\widehat{\mathcal D} = \begin{pmatrix} \mathcal{D} & (0,0)^\top  \\ 
 (0,0) & D_c \end{pmatrix},} 
 \end{align}
 \review{where $\mathcal{D}$ is as in \eqref{diff_matrix_switching} and the drift matrix is}
 \begin{align}\label{drift_matrix_chemo}
 \eqreview{\widehat{\mathcal F} = \begin{pmatrix} \chi \nabla_{{\bf x}} c   & -  (N-1) \epsilon^d \eta_{\bb} \chi b \nabla_{{\bf x}} c &  {\bf 0} \\ (N-1) \epsilon^d \eta_{\rr} \chi r \nabla_{{\bf x}} c   & {\bf 0} &  {\bf 0} \\
 {\bf 0} & {\bf 0} & {\bf 0} \end{pmatrix}.}
 \end{align}
\end{subequations}
\review{As before, this model includes the excluded-volume effects up to $O(\epsilon^2)$ in the equations for $b$ and $r$. Regarding  the chemical, the error in the reduction from the kernel $K_h$ in the equation for $C$ to the source term $\gamma N r$ in the equation for $c$ is $O(h^2)$ (this can be seen using, for example,  Laplace's method). Hence our choice of $h$ ensures that this error is smaller than $O(\epsilon^2)$.}

\review{We conclude this Subsection by performing numerical simulations of the hybrid model for chemotaxis \eqref{sde_chemo} and the reaction-diffusion PDE system \eqref{model_chemo}. Since a key feature of this model is the chemotactic drift, we must adapt our stochastic simulations to account for the drift in the SDEs of blue particles \eqref{SDE_Chemob}. Since Smoldyn \cite{smoldyn} does not presently allow for biased Brownian motion, we perform the particle-based simulations in this Subsection using the alternative library Aboria \cite{Robinson:2017vxa,aboria}. The SDEs \eqref{SDE_Chemob} and \eqref{SDE_Chemor} are again integrated using the Euler--Maruyama method with a constant timestep $\Delta t$. The same timestep is used to integrate the reaction-diffusion PDE \eqref{pde_c}, using the Euler method in time and a second-order finite-difference scheme in space, and to simulate the reactions \eqref{Chemo_reactions}. For details on the implementation of the hybrid model with Aboria we refer the readers to \cite{Bruna:2018ui}.}

\def \scc {0.8}
\def \scl {1.0}
\begin{figure*}
\unitlength=1cm
\begin{center}
\vspace{3mm}
\psfrag{x}[][][\scl]{Position $x$} \psfrag{t}[][][\scl]{Time $t$} \psfrag{tt}[][][\scl]{$t$} \psfrag{dr}[][][\scl]{Density $r$} \psfrag{db}[][][\scl]{Density $b$} \psfrag{dc}[][][\scl]{Concentration $c$} \psfrag{num}[b][][\scl]{Population Number}
\psfrag{a}[][][\scl]{(a)} \psfrag{b}[][][\scl]{(b)} \psfrag{c}[][][\scl]{(c)} \psfrag{d}[][][\scl]{(d)}
\psfrag{Nb}[][][\scl]{$N_{\bb}$} \psfrag{Nr}[][][\scl]{$N_{\rr}$} \psfrag{Nc}[][][\scl]{$N_c$}
\includegraphics[height=.7\columnwidth]{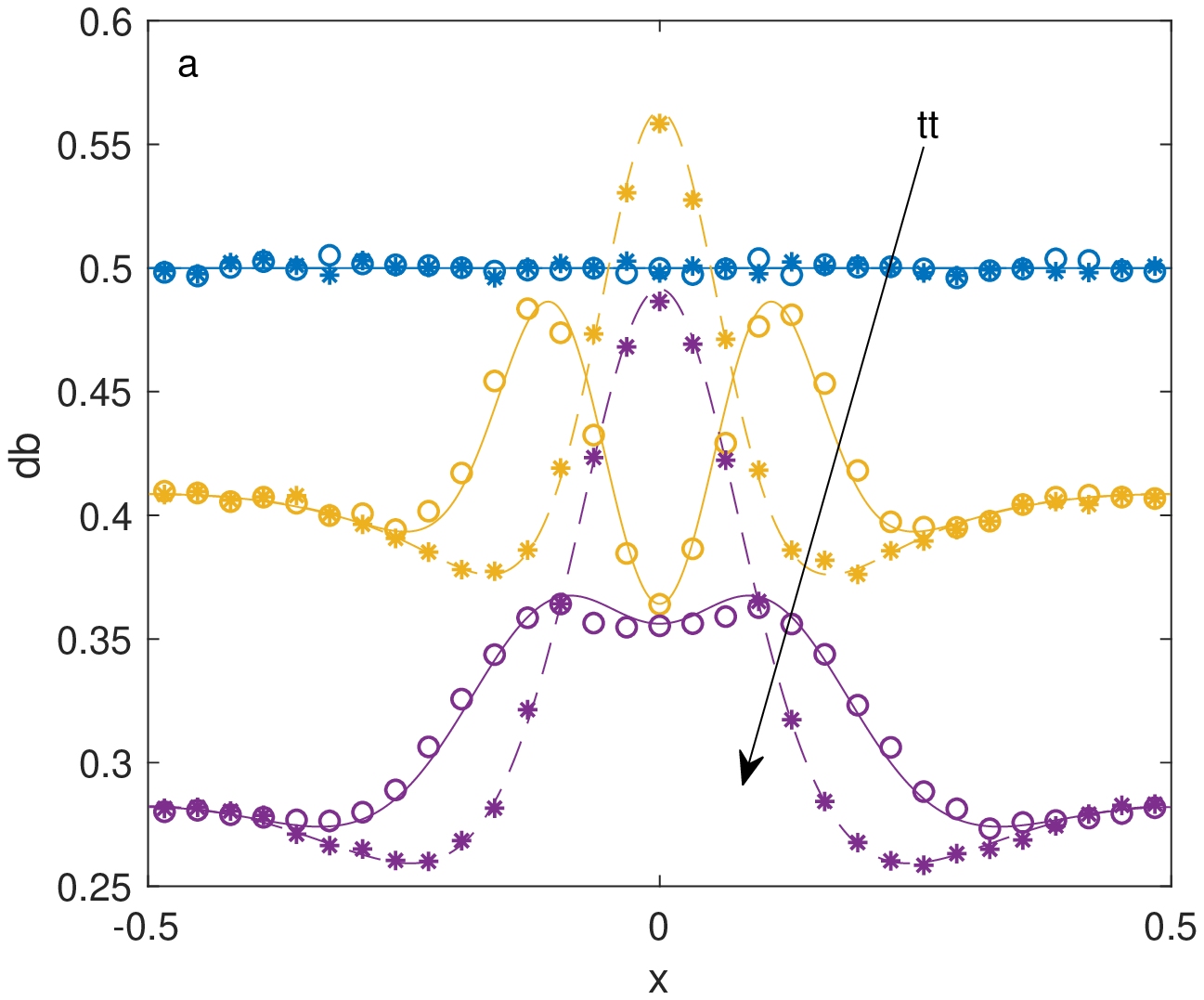} \qquad \includegraphics[height=.7\columnwidth]{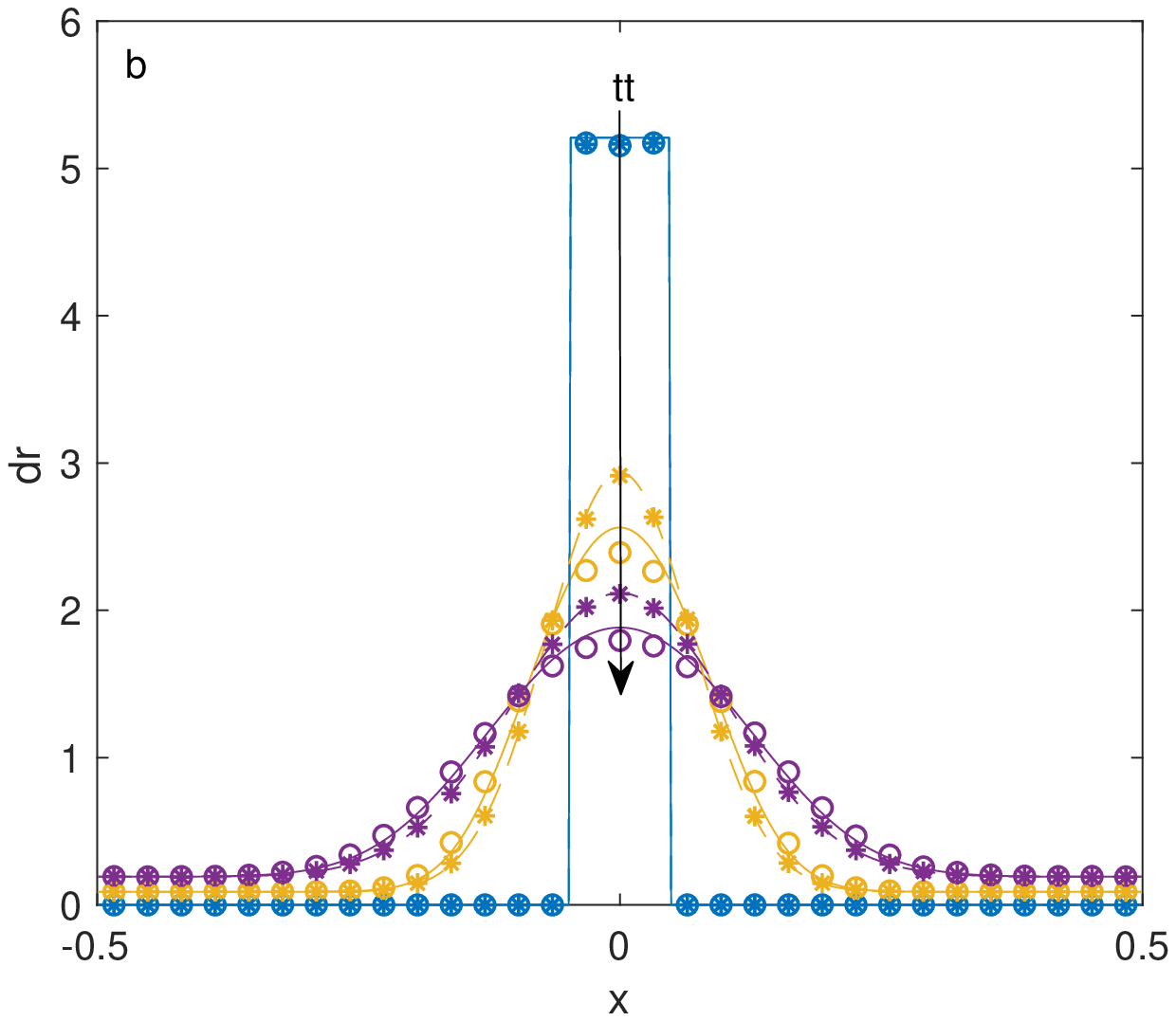} \\ \vspace{.5cm}
\includegraphics[height=.7\columnwidth]{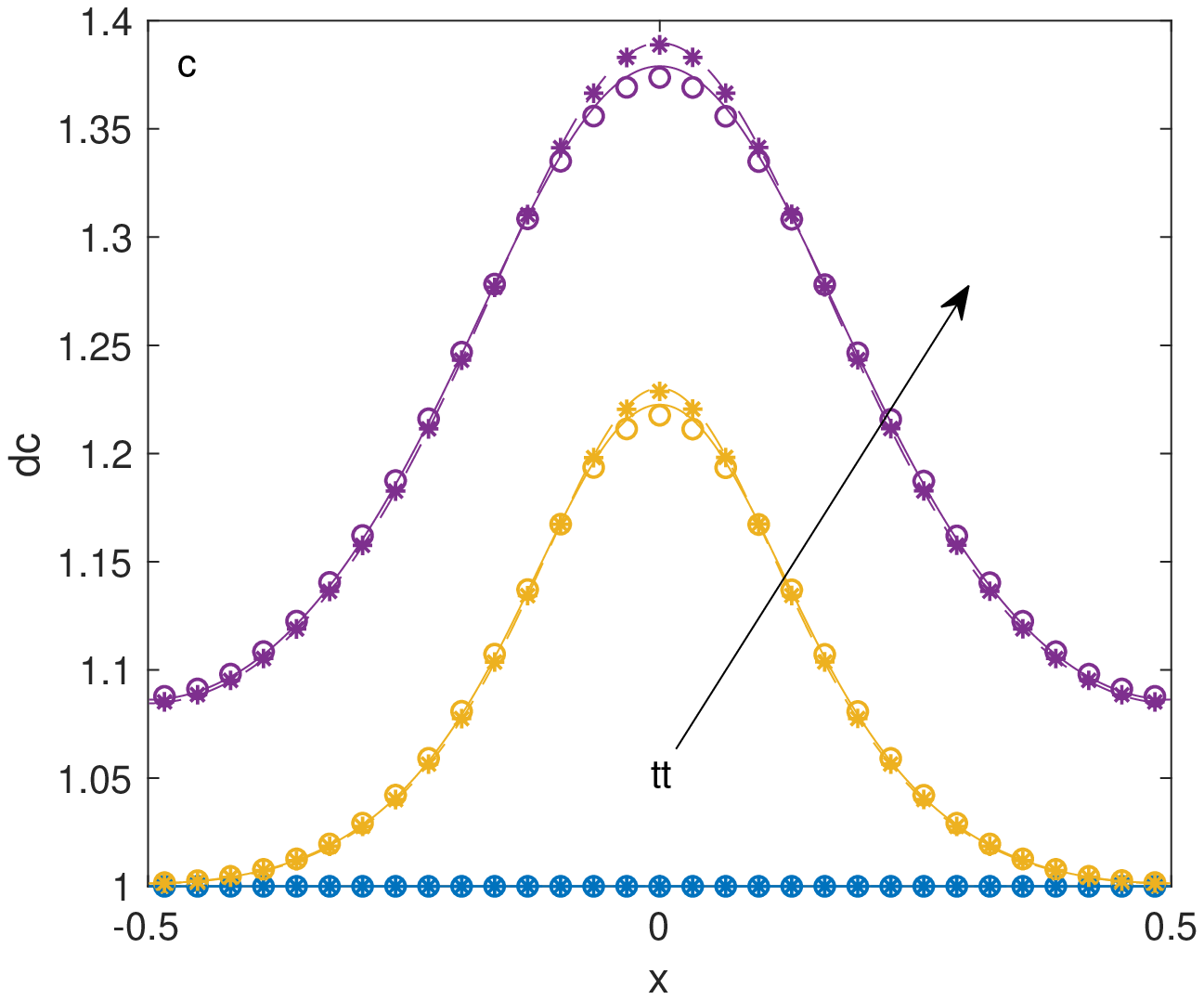} \qquad \includegraphics[height=.7\columnwidth]{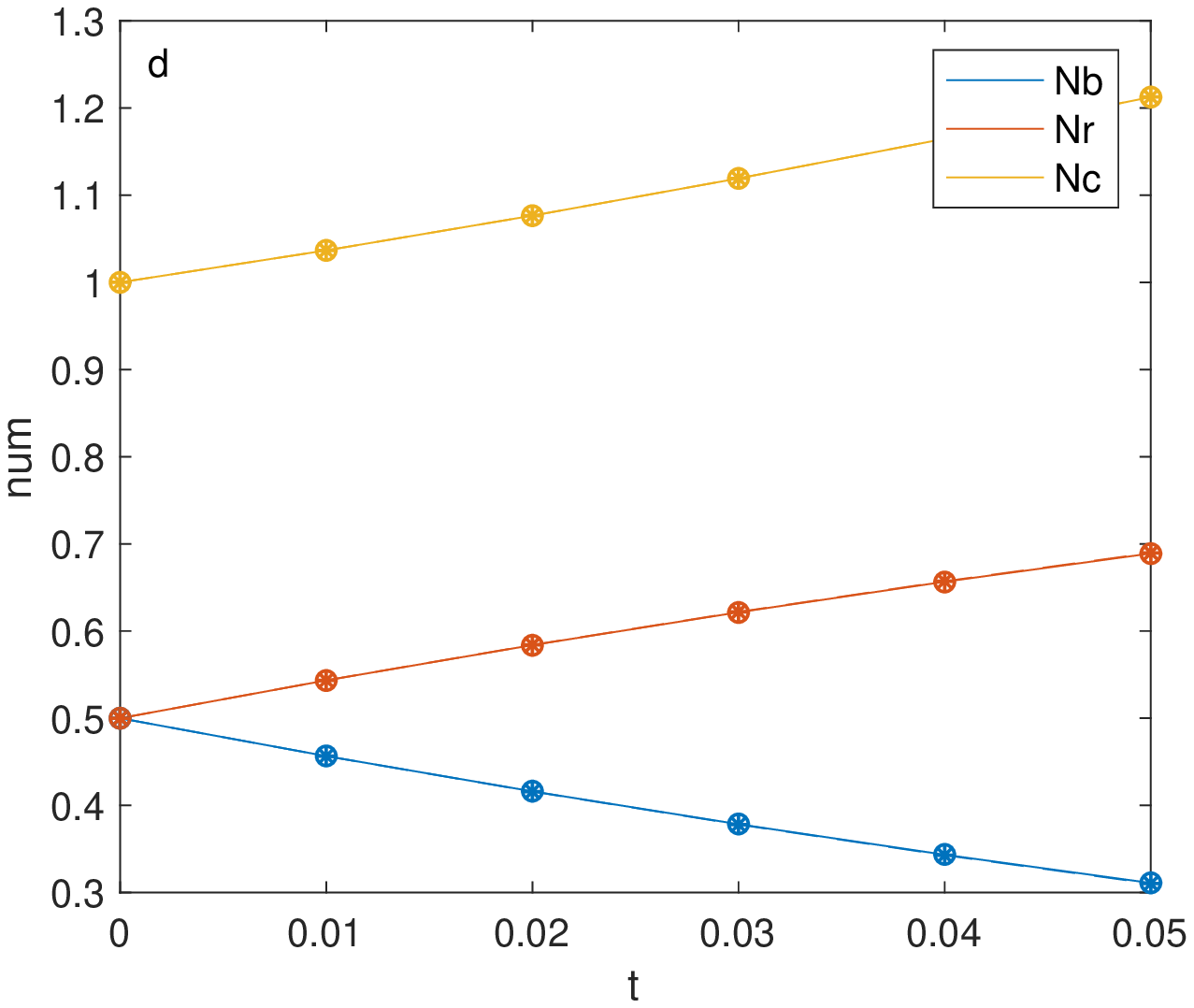}
\caption{\review{Comparison between the stochastic hybrid chemotaxis model \eqref{sde_chemo} (asterisks for $\epsilon = 0$ and circles for $\epsilon = 0.01$) and continuum PDE model \eqref{model_chemo} (dashed lines for $\epsilon = 0$ and solid lines for $\epsilon = 0.01$) for a system with $N=800$ cells with $N_{\bb}(0) = N_{\rr}(0) = 400$ and final time $T_f = 0.05$. (a) Evolution of the blue cells density $b(\bfx, t)$, with uniform initial density. (b) Evolution of the red cells density $r(\bfx, t)$, with initial density uniform in $[-3/62, 3/62]\times [-1/2,1/2]$. (c) Evolution of the chemoattractant concentration $c(\bfx, t)$ with initial concentration $c_0(\bfx) = 1$. 
(d) Evolution of the cell numbers $N_{\bb}(t)$ and $N_{\rr}(t)$ and the total concentration $N_c = \int_\Omega c(\bfx, t) \ud \bfx$. 
Figures (a), (b), (c) show the solutions averaged over the vertical direction at times $t = 0, 0.01, 0.02, 0.05$. 
Other parameters used: $D_{\bb} = D_{\rr} = 0.1$, $D_c = 1$, $\chi = 1$, $k_\bb = 10$, $k_\rr = 1$, $\gamma = 0.01$, $\mu = 0.5$. Histograms computed from 5000 realizations and 31 bins in the horizontal coordinate. PDE model solved using 500 grid points.}}
\label{fig:fig3}
\end{center}
\end{figure*}

\review{Figure \ref{fig:fig3} shows a comparison between the hybrid model \eqref{sde_chemo} and continuum model \eqref{model_chemo} of chemotaxis with and without excluded-volume interactions. We start with a population of $N=800$ cells of diameter $\epsilon=0.01$ in the domain $\Omega = [-1/2,1/2]^2$. Half of the particles are initially blue and uniformly distributed, and the other half are red and uniformly distributed in a vertical band centered at $x=0$ of width 3/31. The initial condition for the chemical is $c_0(x) = 1$. We perform 5000 realizations of the stochastic simulation so that, as before, we have $4\times 10^6$ trajectories. Figures \ref{fig:fig3}(a-c) show the densities $b$, $r$, and  the concentration $c$ averaged over the vertical direction at four different times $t$. As time progresses, the red particles produce chemoattractant and this leads to a higher concentration of chemical in the centre of the domain (see Figure \ref{fig:fig3}(c)). This gradient in concentration induces a drift to the blue particles, that want to move towards the middle of the domain. In the case of point particles, this gives rise to a hill-shaped profile of the blue density centered at the origin. However, in the case of finite-size particles, the interplay between the gradient of concentration $\nabla_\bfx c$ and the gradient of reds $\nabla_\bfx r$ results in a volcano shaped profile of the blue density (see Figure \ref{fig:fig3}(a)). There is a net production of red particles to the detriment of the blue particles and a net decay of chemical $c$ (Figure \ref{fig:fig3}(d)). We note that, as for the example in Figure \ref{fig:fig2}(b), the time evolution of the total number of blue and red particles is not affected by the excluded-volume interactions.}
\review{We find good agreement between the hybrid and continuum models. The relative $L_2$-norm errors in the densities in Figures \ref{fig:fig3}(a-c) (using the solutions at the histogram grid points and $t = 0, 0.01, 0.02, \dots, 0.05$) are, for point particles ($\epsilon = 0$), $0.39\%, 0.85\%$, and $0.038\%$, respectively. The errors in the case of $\epsilon = 0.01$ are $0.80\%, 4.2\%$, and $0.15\%$, respectively.  We note that these errors are very small compared to errors we would have committed by ignoring the excluded-volume effects: the errors between the PDE model for $\epsilon= 0$ and the stochastic model for $\epsilon=0.01$ are $12\%, 13\%$ and $0.40\%$ for the blue particles, red particles, and chemical concentration, respectively.}

\section{Switching Upon Collision: Bimolecular reactions} \label{Coll}

In this section we assume that particles switch between subpopulations as a result of a pairwise collision or interaction and consider bimolecular reactions of the form $B + R \rightarrow 2R$. We first describe how this effect changes our particle-level description and then derive the corresponding population-level equations.

\subsection{Particle-based model}

Our discrete model has the same Langevin dynamics as before  \eqref{sde_switch}. When two blue (red) particles are in contact, \review{$\| {\bf X}_i-{\bf X}_j\| = \epsilon $ and $i,j \in \mathcal I_{\bb}(t)$ or $\mathcal I_{\rr}(t)$},  we impose standard reflective boundary conditions. However, when a blue and a red particle are in contact, \review{$i \in \mathcal I_{\bb}(t)$ and $j \in \mathcal I_{\rr}(t)$, the blue particle becomes red with a certain probability, resulting in $S_i(t) = \rr$ and $N_{\bb}(t) = N_{\bb}(t) - 1$ after the collision. }

As before the joint probability density function $P(\vec{x},\vec{s},t)$ obeys the linear Fokker--Planck equation
\begin{align} \label{BS_PPDE}
\partial_t P  = \nabla_{\vec{x}} \cdot \left[ {\mathsf D}_{\vec{s}} \nabla_{\vec{x}} P(\vec{x}, \vec{s},t) - \vec{F}_s(\vec{x}, \vec s) P(\vec{x}, \vec{s},t) \right],
\end{align}
\review{for $\vec{x} \in \Omega_{\epsilon}^N$}. The boundary conditions  now require more careful treatment. We define a flux for each configuration of particle states
\begin{align}
\mathcal{J}_{\vec{s}} = \left[ {\mathsf D}_{\vec{s}} \nabla_{\vec{x}} P(\vec{x}, \vec{s},t) - \vec{F}(\vec{x}, \vec s) P(\vec{x}, \vec{s},t) \right] \cdot \vec{n}.
\end{align}
In the case of spontaneous switching on the internal boundaries $\partial \Omega_{\epsilon}^N \setminus \partial \Omega^N $ we had that $\mathcal{J}_{\vec{s}}=0$ for all $\vec{s} \in \mathcal{C}^{N}$. However there are now non-zero fluxes across these internal boundaries in the configuration space, where the reactivity parameter for these boundaries is $\lambda$. To make this explicit, we consider the boundary between a pair of particles, $\|{\bf x}_i - {\bf x}_j \| = \epsilon$ for $i \neq j$. If ($s_i,s_j$)=($\bb,\bb$) then $\mathcal{J}_{\vec{s}} = 0$ on this boundary, as there is no reaction between two blue particles. If ($s_i,s_j$)=($\bb,\rr$) or ($\rr,\bb$), then we have $\mathcal{J}_{\vec{s}} = \lambda P_{\vec{s}}$ on $\|{\bf x}_i - {\bf x}_j \| = \epsilon$. This is because there is a flux out of the configuration state $\vec{s}$ due to the bimolecular reaction. Finally if ($s_i,s_j$)=($\rr,\rr$) no reaction occurs for two red particles; however there is a flux into the configuration state $\vec{s}$ on the boundary $\|{\bf x}_i - {\bf x}_j \| = \epsilon$ from the blue and red reactions. Therefore on this boundary $\mathcal{J}_{\vec{s}} = - \lambda \left( P_{\vec{c}_i} + P_{\vec{c}_j} \right)$, where $\vec{c}_i = (s_1,\ldots,s_{i-1},\bb,s_{i+1},\ldots,s_N)$ and $\vec{c}_j = (s_1,\ldots,s_{j-1},\bb,s_{j+1},\ldots,s_N)$.

\subsection{Population-based model}

To reduce the particle-level description to the population-level we consider as before the density functions \eqref{br_densities}. We restrict attention to the case for $N=2$ and \review{rewrite the problem in terms of $P_{s_1s_2}({\bf x}_1,{\bf x}_2,t) = P({\bf x}_1,{\bf x}_2,s_1,s_2,t)$ as before.} For simplicity, we again set ${\bf f}_\bb = {\bf f}_\rr = {\bf 0}$. For $N=2$ we only have one inner boundary, $\|{\bf x}_1 - {\bf x}_2 \| = \epsilon$, with the following boundary conditions
\begin{align}
\mathcal{J}_{\bb \bb} &= D_{\bb} \nabla_{{\bf x}_1} P_{\bb \bb} \cdot \mathbf{n_1} + D_{\bb} \nabla_{{\bf x}_2} P_{\bb \bb} \cdot \mathbf{n_2} = 0,   \nonumber \\
\mathcal{J}_{\bb \rr} &= D_{\bb} \nabla_{{\bf x}_1} P_{\bb \rr} \cdot \mathbf{n_1} + D_{\rr} \nabla_{{\bf x}_2} P_{\bb \rr} \cdot \mathbf{n_2} = \lambda P_{\bb \rr},\label{J_br}\\
\mathcal{J}_{\rr \bb} &= D_{\rr} \nabla_{{\bf x}_1} P_{\rr \bb} \cdot \mathbf{n_1} + D_{\bb} \nabla_{{\bf x}_2} P_{\rr \bb} \cdot \mathbf{n_2} = \lambda P_{\rr \bb},  \nonumber\\
\mathcal{J}_{\rr \rr} &= D_{\rr} \nabla_{{\bf x}_1} P_{\rr \rr} \cdot \mathbf{n_1} + D_{\rr} \nabla_{{\bf x}_2} P_{\rr \rr} \cdot \mathbf{n_2} = -\lambda(P_{\bb \rr}\!+\!P_{\rr \bb}).  \nonumber
\end{align}
Then, from (\ref{BS_PPDE}) the equations for the blue and red densities with the collision integrals can be written as
\begin{subequations} \label{br_collision}
\begin{align}
\partial_t b =& D_{\bb} \nabla_{{\bf x}_1}^2 b  + \! \int_{\partial B_{\epsilon}({\bf x}_1)}\! -\big[ 2D_{\bb} \nabla_{{\bf x}_1} (P_{\bb \bb}+P_{\bb \rr} )  \label{BS_CI_b} \\
& \hspace{2.8cm} + (D_{\bb}\!-\!D_{\rr}) \nabla_{{\bf x}_2} P_{\bb \rr} \big] \!\cdot \mathbf{n_2} \, \mathrm{d}S_{{\bf x}_2}, \nonumber \\
\partial_t r  =& D_{\rr} \nabla_{{\bf x}_1}^2 r  + \!\int_{\partial B_{\epsilon}({\bf x}_1)}\! -\big[ 2D_{\rr} \nabla_{{\bf x}_1} (P_{\rr \bb}+P_{\rr \rr} )  \label{BS_CI_r} \\
& \hspace{2.8cm} + (D_{\rr}\!-\!D_{\bb}) \nabla_{{\bf x}_2} P_{\rr \bb} \big] \! \cdot \mathbf{n_2} \,  \mathrm{d}S_{{\bf x}_2}.  \nonumber
\end{align}
\end{subequations}
We next compute the asymptotic expansions for $P_{\bb \bb}$, $P_{\bb \rr}$, $P_{\rr \bb}$ and $P_{\rr \rr}$ in the inner region in order to evaluate the  integrals in \eqref{br_collision}. 

\subsection{Matched asymptotic expansions and collision integral}

First we consider $P_{\bb \bb}({\bf x}_1,{\bf x}_2,t)$. As the boundary condition on $\|{\bf x}_1 - {\bf x}_2 \| = \epsilon$ is $\mathcal{J}_{\bb \bb}=0$, we can use the results of Section \ref{Coll_C} to write down the inner solution as 
\begin{equation} \label{Pbb_collision}
\hat{P}_{\bb \bb}(\hat{{\bf x}}_1, \hat{{\bf x}}, t ) \sim q_{\bb}^2(\hat{{\bf x}}_1,t) + \epsilon q_{\bb}(\hat{{\bf x}}_1,t) \hat{{\bf x}} \cdot \nabla_{\hat{{\bf x}}_1} q_{\bb}(\hat{{\bf x}}_1,t) ,
 \end{equation}
for some function $q_{\bb}$. 

We construct the inner solution to $P_{\bb \rr}({\bf x}_1,{\bf x}_2,t)$ by matching the solutions in the outer and inner regions. Introducing the inner variables ${\bf x}_1=\hat{{\bf x}}_1$ and ${\bf x}_2=\hat{{\bf x}}_1+\epsilon \hat{{\bf x}}$ as before,
we transform equations (\ref{BS_PPDE}) and (\ref{J_br}) to get the following problem
\begin{subequations} \label{inner_switch}
\begin{multline}
\epsilon^2 \partial_t \hat{P}_{\bb \rr} = (D_{\bb}+D_{\rr})\nabla_{\hat{{\bf x}}}^2 \hat{P}_{\bb \rr} \\ - 2\epsilon D_{\bb} \nabla_{\hat{{\bf x}}_1} \cdot \nabla_{\hat{{\bf x}}} \hat{P}_{\bb \rr} + \epsilon^2  D_{\bb} \nabla_{\hat{{\bf x}}_1}^2 \hat{P}_{\bb \rr}, \label{BS_inner_PDE}
\end{multline} 
with boundary conditions
\begin{equation}
\hat{{\bf x}} \cdot \nabla_{\hat{{\bf x}}} \hat{P}_{\bb \rr} = \frac{\epsilon}{D_{\bb} + D_{\rr}}( D_{\bb}\hat{{\bf x}} \cdot \nabla_{\hat{{\bf x}}_1} \hat{P}_{\bb \rr} + \lambda \hat{P}_{\bb \rr} ),  \label{BS_innek_bC}
\end{equation}
\end{subequations}
\review{on $\|{\bf x}\| = 1$, and matching condition given by \eqref{matching_switch} as $\hat \bfx \to \infty$.} We seek a regular power series expansion of the form $\hat{P}_{\bb \rr} = \hat{P}_{\bb \rr}^{(0)} + \epsilon \hat{P}_{\bb \rr}^{(1)} + \cdots$, and as in Section \ref{Spont} the zeroth order problem has the trivial solution $\hat{P}_{\bb \rr}^{(0)} = q_{\bb}(\hat{{\bf x}}_1,t)q_{\rr}(\hat{{\bf x}}_1,t)$. From \eqref{inner_switch}, the problem for $\hat{P}_{\bb \rr}^{(1)}$ is
\begin{subequations} \label{o1_collision}
\begin{alignat}{2}
\nabla_{\hat{{\bf x}}}^2 \hat{P}_{\bb \rr}^{(1)} &= 0,  \\
\hat{{\bf x}} \cdot \nabla_{\hat{{\bf x}}} \hat{P}_{\bb \rr}^{(1)} & = \dfrac{D_{\bb}}{D_{\bb}+D_{\rr}} \hat{{\bf x}} \cdot \nabla_{\hat{{\bf x}}_1} (q_{\bb} q_{\rr}) \nonumber \\
& \quad + \dfrac{\lambda}{D_{\bb}+D_{\rr}} q_{\bb}q_{\rr},  \quad \text{on} \  \|\hat{{\bf x}}\|=1, \\
\hat{P}_{\bb \rr}^{(1)} & \sim \hat{{\bf x}} \cdot q_{\bb}\nabla_{\hat{{\bf x}}_1} q_{\rr}, \quad \text{as}  \ \|\hat{{\bf x}}\| \rightarrow \infty. 
\end{alignat}
\end{subequations}

In two dimensions ($d=2$), problem \eqref{o1_collision} does not have a solution satisfying both the inner and outer boundary conditions, indicating that a more intricate method is required.  For the case $d=3$ the solution is
\begin{align*}
\hat{P}_{\bb \rr}^{(1)} = \ &   q_{\bb} \hat{{\bf x}} \cdot \nabla_{\hat{{\bf x}}_1} q_{\rr}  - \dfrac{\lambda}{(D_{\bb}+D_{\rr}) \|\hat{{\bf x}}\| } q_{\bb} q_{\rr}   \\
& + \dfrac{\hat{{\bf x}}}{2(D_{\bb}+D_{\rr})\|\hat{{\bf x}}\|^3} ( D_{\rr} q_{\bb} \nabla_{\hat{{\bf x}}_1} q_{\rr} - D_{\bb} q_{\rr} \nabla_{\hat{{\bf x}}_1} q_{\bb} ).
\end{align*}
Combining the results above we find that 
\begin{align} \label{Pbr_collision}
\hat{P}_{\bb \rr} \sim   q_{\bb}q_{\rr} &+ \epsilon \bigg[ q_{\bb} \hat{{\bf x}} \cdot \nabla_{\hat{{\bf x}}_1} q_{\rr} -\dfrac{\lambda}{(D_{\bb}+D_{\rr})\|\hat{{\bf x}}\| } q_{\bb}q_{\rr} \\
&+ \dfrac{\hat{{\bf x}}}{2(D_{\bb}+D_{\rr})\|\hat{{\bf x}}\|^3} ( D_{\rr} q_{\bb} \nabla_{\hat{{\bf x}}_1} q_{\rr}  - D_{\bb} q_{\rr} \nabla_{\hat{{\bf x}}_1} q_{\bb} \bigg ]. \nonumber
\end{align}
We denote by $I$ the collision integral in equation (\ref{BS_CI_b}),
\begin{multline*}
I =  \int_{\partial B_{\epsilon}({\bf x}_1)} - [ 2D_{\bb} \nabla_{{\bf x}_1}  (P_{\bb \bb}+P_{\bb \rr} ) \\ + (D_{\bb}- D_{\rr}) \nabla_{{\bf x}_2} P_{\bb \rr} ] \cdot \mathbf{n_2} \mathrm{d}S_{{\bf x}_2}.
\end{multline*}
Using the inner solutions \eqref{Pbb_collision} and \eqref{Pbr_collision} we can now evaluate $I$. Expressing it in terms of inner variables and using the boundary condition (\ref{BS_innek_bC}), we obtain
\begin{align}
\label{I_collision}
\begin{aligned}
I \sim &- 4\pi \epsilon^2 \lambda  q_{\bb} q_{\rr}  + 4 \pi\epsilon^3 \dfrac{\lambda^2}{D_{\bb}+D_{\rr}}  q_{\bb} q_{\rr} \\
&+ \epsilon^3 D_{\bb} (  \omega  q_{\bb} \nabla_{{{\bf x}}_1} q_{\bb} +
 \xi_{\bb} q_{\bb} \nabla_{{{\bf x}}_1} q_{\rr}  
	 - \eta_{\bb} q_{\rr} \nabla_{\hat{{\bf x}}_1} q_{\bb} ),
\end{aligned}
\end{align}
where the constants $\omega$, $\xi_{\bb}$ and $\eta_{\bb}$ are defined in (\ref{parameters}). We use again the normalization condition to replace $q_{\bb}$ and $q_{\rr}$ by $b$ and $r$, respectively. Combining these results we deduce that the population-level equation for a blue particles $b(\bfx, t)$ satisfies
\begin{multline}
\partial_t b = D_{\bb} \nabla_{{\bf x}} \cdot \big[ ( 1 + \omega \epsilon^3 b ) \nabla_{{\bf x}} b + \epsilon^3  ( \xi_{\bb} b \nabla_{{\bf x}} r - \eta_{\bb} r \nabla_{{\bf x}} b ) \big] \\
-4\pi\epsilon^2 \lambda \left( 1 - \epsilon \dfrac{\lambda}{D_{\bb}+D_{\rr}} \right) br,
\end{multline}
From (\ref{BS_CI_r}) a similar equation can be derived for the red density. For the general case of $N$ particles, reintroducing drift terms, we have the following system of nonlinear reaction-diffusion-advection equations
\begin{align} \label{BS_populationlevel}
\partial_t \begin{pmatrix} b \\ r \end{pmatrix} &=  \nabla_{{\bf x}} \cdot \left[ {\mathcal D} \,  \nabla_{{\bf x}}  \begin{pmatrix} b \\ r \end{pmatrix} - {\mathcal F} \begin{pmatrix} b \\ r \end{pmatrix}
 \right] + \mathcal{R},
\end{align}
\review{for ${\bf x} \in \Omega$,} where the diffusion matrix, $\mathcal{D}$, and the drift matrix, ${\mathcal F}$, are defined by equations (\ref{diff_matrix_switching}) and (\ref{drift_matrix_switching}). The reaction terms $\mathcal{R}$ are given by
\begin{equation} \label{reaction_excluded}
\mathcal{R} = \lambda (N-1) 4 \pi \epsilon^2 \left(1 - \epsilon \dfrac{\lambda}{D_{\bb}+D_{\rr}} \right) br \begin{pmatrix} -1 \\ 1
\end{pmatrix}.
\end{equation}
We interpret the reaction terms as follows: $(N-1)4 \pi \epsilon^2$ is the total surface area of the $(N-1)$ spheres with which the sphere under consideration could react. The factor $\left( 1-\epsilon \lambda / (D_{\bb}+D_{\rr}) \right)$ represents a correction to the source and sink terms, and indicates that only a fraction of the available surface area is used. If the diffusion coefficients are large relative to the reactivity parameter $\lambda$ then this correction term approaches one, which can be interpreted as the case when the particle is moving fast enough that the entire surface area is available to it for collisions. 

Our result \eqref{reaction_excluded} can be related to the $\lambda$ - $\bar \rho$ model introduced by Erban and Chapman~\cite{erban2009stochastic} as follows. In their model, a bimolecular reaction can only take place when the distance between two particles is less than $\bar \rho$, and it occurs at a rate $\lambda$. They distinguish between two cases depending on whether the relative mean-square displacement between the particles in one timestep, $\sqrt{2(D_{\bb} + D_{\rr}) \Delta t}$, is small or large compared to the reaction radius $\bar \rho$. We consider the former case, since we already perform simulations in this regime so that we do not miss collisions between particles (as discussed in Section \ref{CollNR}). Taking the limit $\lambda$ small, the reaction rate was found to be proportional to the volume of the reactive region (see Eq. (31) in \cite{erban2009stochastic}). By contrast, the leading order of our reaction rate is proportional to the surface area of the collision surface (see \eqref{reaction_excluded}). This discrepancy arises because we have incorporated the excluded-volume effect directly into our microscopic model and, further, particles are not allowed to get closer to each other than the reaction radius, that is, $\bar \rho \equiv \epsilon$ in our case. 

\subsection{Numerical Results}

In this section we show how the inclusion of bimolecular reactions affects the time dependent behavior of our model. The stochastic simulations are implemented as described previously in Section \ref{CollNR} with the exception of the reaction terms. When considering particle switching due to the collisions we proceed as follows. When a red and blue particle overlap, the reaction is implemented by generating a random number $\rho$ uniformly on the interval $[0,1]$. If $\rho < \mathcal{P} \sqrt{\Delta t}$, where $\mathcal{P} \sqrt{\Delta t}$ is the reaction probability, we switch the color of the blue particle to red and then update the positions due to the collision. We note that the relationship between the reactivity parameter, $\lambda$, and the reaction probability parameter, $\mathcal{P}$, has been shown to be \cite{erban2007BCs}
\begin{equation}
\mathcal{P}  = \dfrac{\lambda \sqrt{\pi}}{\sqrt{D_{\bb}+D_{\rr}}}.
\end{equation}
In order to increase the accuracy of this method we may also want to account for the possibility, that during the interval $[t,t+\Delta t]$, the red and blue particles may have overlapped and then moved apart, and a reaction may have occurred that otherwise would be missed. This correction, first pointed out by Andrews and Bray \cite{andrews2004stochastic}, is not included in our implementation of the bimolecular reactions. Instead, the timestep $\Delta t$ is taken to be small enough that the probability of particles overlapping and separating during $\Delta t$ is negligible \review{\cite{andrews2004stochastic}}.

\begin{figure}
\begin{center}
	\includegraphics[width=.45\textwidth]{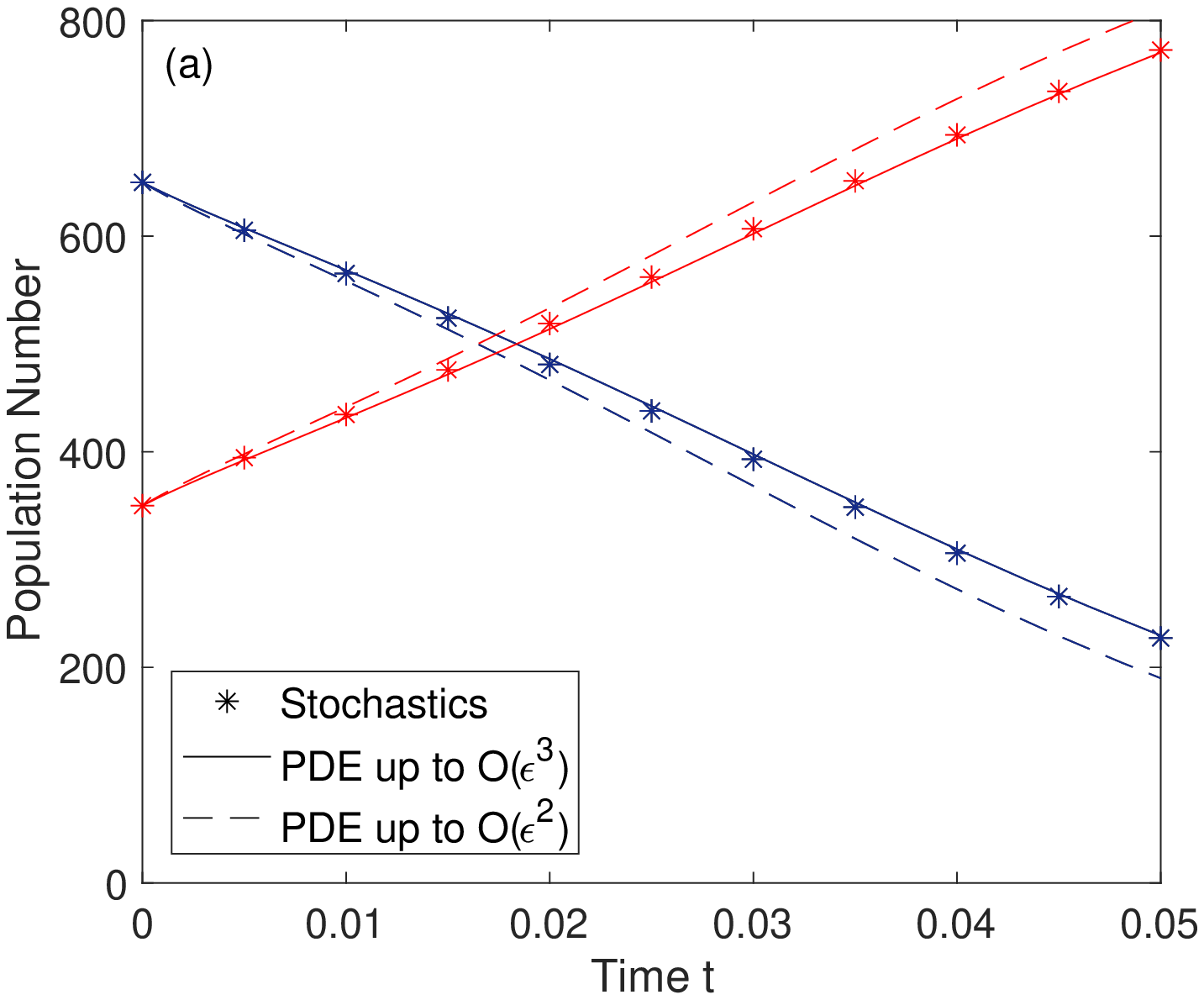} \quad \includegraphics[width=.45\textwidth]{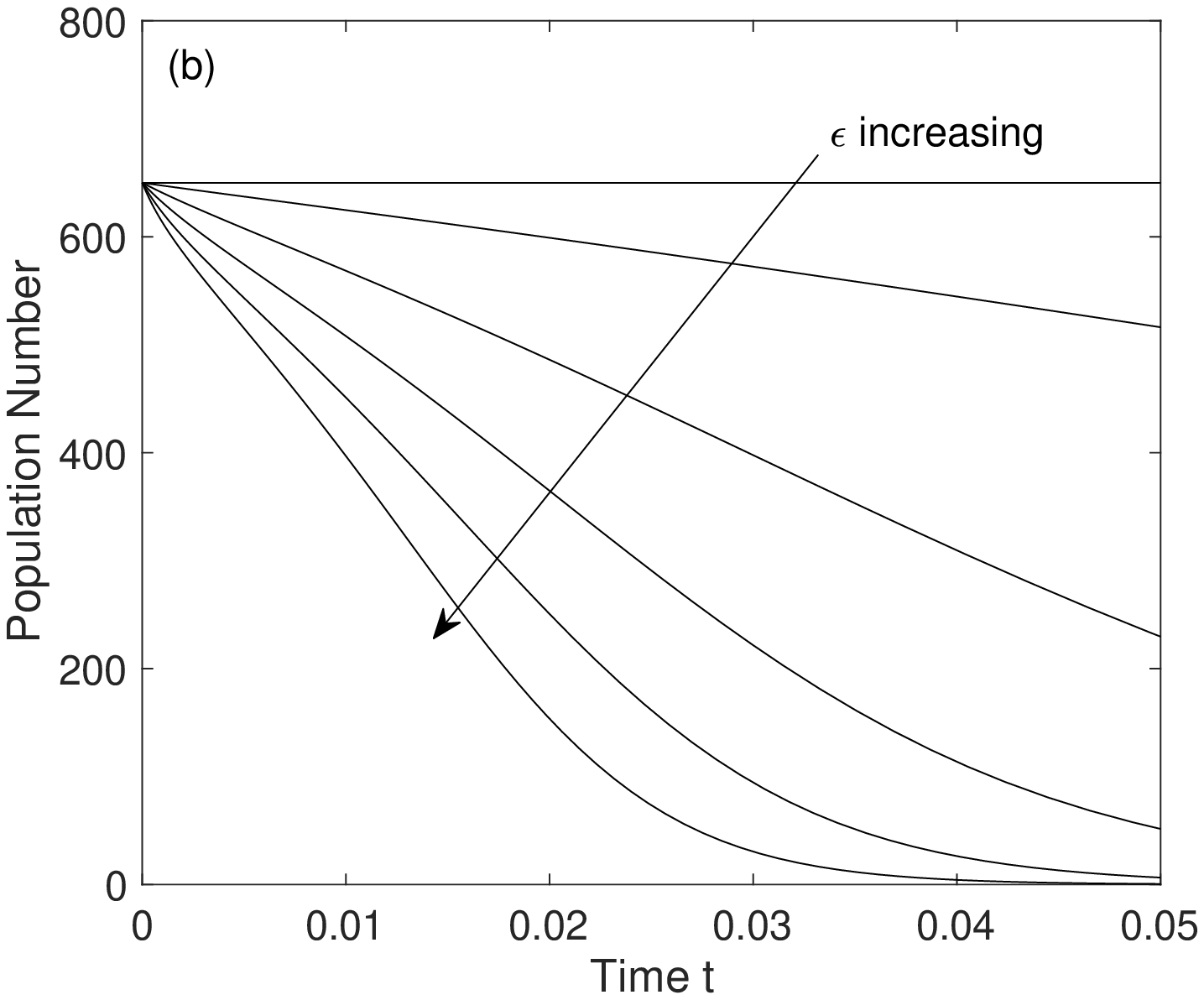}
\end{center}
\caption{ a) Plots of the numerical solutions to equation (\ref{BS_integro}) correct to  $O(\epsilon^2)$ and  $O(\epsilon^3)$, where $\epsilon = 0.02$.   b) Plots of the evolution of the blue population number, $N_{\bb}(t)$, for a variety of diameters $\epsilon=\{0.00,0.01,0.02,0.03,0.04,0.05\}$. Parameter values: $D_{\bb}=0.5$, $D_{\rr}=1$, $\lambda=10$, ${\bf f}_\bb = {\bf f}_\rr = {\bf 0}$ and $N=1000$. The initial conditions were $N_{\bb}(0)=650$ and the particles were uniformly distributed in $\Omega = [-1/2,1/2]^3$ and $N_{\rr}(0)=350$ which were normally distributed with zero mean and standard deviation $0.09$ in the $(x,y)$ plane and uniformly distributed in the $z$ coordinate.}
\label{BS_PopNum}
\end{figure}

We consider the evolution of the population numbers by integrating equations (\ref{BS_populationlevel}) over the domain $\Omega$ to obtain the following integro-differential equation for the number of blue particles $N_{\bb}(t)$
\begin{align}\label{BS_integro}
N_{\bb}' = - 4 \pi \lambda \epsilon^2  N(N-1) \bigg(1 - \epsilon\dfrac{\lambda}{D_{\bb}+D_{\rr}} \bigg) \!\int_{\Omega}  \! b r \, \mathrm{d}{\bf x},
\end{align}
and $N_{\rr}(t) = N - N_{\bb}(t)$. Since the leading-order reaction term in equation (\ref{BS_integro}) is at order $\epsilon^2$, it is natural to ask if retaining terms to $O(\epsilon^2)$ provides sufficient detail to accurately predict the evolution of the population numbers. In Figure \ref{BS_PopNum}(a) we present the solutions to equation (\ref{BS_integro}) correct to $O(\epsilon^2)$ and  $O(\epsilon^3)$. We note that the $O(\epsilon^3)$ terms are needed to accurately model the evolution of the population number. \review{In particular, the relative $L_2$-norm errors of the PDE model with only the $O(\epsilon^2)$ reaction terms are $5.9\%$ and $4.0\%$ for blue and red particles respectively. The relative errors when adding the $O(\epsilon^3)$ reactions terms  in the PDE are reduced to $0.9\%$ and $0.6\%$, respectively.}

We also consider how the total volume fraction occupied by spheres affects the evolution of the population of particles. To investigate the effect of volume fraction we plotted the evolution of the blue population for a variety of different diameters $\epsilon$ in Figure \ref{BS_PopNum}(b), keeping $N$ fixed. We see that the time to extinction decreases as volume fraction increases. In particular, when only $6.55\%$ of the domain is filled with particles we observe rapid extinction of the blue particles population.

\section{Discussion} \label{Discuss}

\review{In this paper we have studied the interplay between volume exclusion and reactions in a two-species system of Brownian hard-sphere particles. Reactions between the two subpopulations of particles, which we termed as either ``blue" or ``red" particles, are either spontaneous or as a result of hard-core collisions.}

In Section \ref{Spont} we considered the case of spontaneous switching. We formulated the discrete model, a system of overdamped Langevin SDEs, to allow particles to switch their color at a spatially dependent rate. The resulting system of nonlinear reaction-diffusion-advection equations \eqref{model_spo} is similar to that for multiple populations without switching \cite{bruna2012multiple} with additional linear source and sink terms. We then presented numerical solutions of the PDEs and showed that they are in good agreement with stochastic simulations of the discrete model.
\review{We also showed how our modeling framework can be adapted to a well-studied problem in mathematical biology, namely that of cell chemotaxis. In this case, the spontaneous reaction from blue to red particles is mediated by the chemical. By introducing a third species of point particles representing the chemoattractant, we showed how to coarse-grain a hybrid model whereby cells are modeled as individual particles of a finite-size and the chemical as a continuum already, to a continuum model for all three species. The hybrid model is similar to that used by \cite{McLennan:2012jb}, but here we were able to show how the hybrid model upscales to a continuum model. In particular, numerical simulations of the hybrid and continuum models showed that excluded-volume interactions in combination with reactions and chemotaxis can have a significant impact on the distribution of cells in the domain.}

In Section \ref{Coll} we considered bimolecular reactions due to collisions between the hard-sphere particles. In particular, we considered the reaction $B + R \rightarrow 2R$, so that a blue particle may change to red, with a prescribed probability, after a collision with a red particle. \review{This bimolecular reaction comes into the model as a nontrivial boundary condition in the high-dimensional Fokker--Planck PDE and, as a result, it changes the problem to solve via matched asymptotic expansions substantially. We obtained a solution for the three-dimensional case. The case of two spatial dimensions is harder, because a random walker will return to its starting point with probability one in two dimensions, but not in three \cite{Durrett}.}
We leave the two-dimensional case, which could be solved by introducing an intermediate region between the inner and outer regions in the method of matched asymptotic expansions, for future work. \review{The cross-diffusion model with bimolecular reactions between subpopulations is the main contribution of this paper. To our knowledge, this is the first continuum PDE model for (off-lattice) Brownian particles that combines diffusion and reactions occurring due to contact between particles, instead of introducing an artificial reaction radius \cite{erban2009stochastic}.}

\review{The method presented in this paper provides a systematic way of linking the stochastic particle-level description and the population-level PDE description. As such it is ideally suited for experimental validation. Depending on the type of experimental data available for a given biological system, either the stochastic model or the PDE model might be more readily parameterized. Then one could exploit the link between the two levels of description to validate the method and learn more about the experimental system. This study is beyond the scope of the present work.}

An interesting extension of this framework would be to incorporate cell proliferation and death. The main \review{challenges} when including reactions of this type into the modeling framework is that the total number of cells in the system, $N(t)$, becomes a random variable \review{and as such the dimensionality of the configuration space can fluctuate}. A further consideration is that of the volume fraction occupied in the domain: unless parent cells divide to produce offspring (keeping the total volume constant), the volume occupied in the domain will increase. 

\begin{acknowledgments}
The authors thank Dr. Steve Andrews and Dr. Martin Robinson for helpful discussions on the implementation of the stochastic algorithms using the software libraries Smoldyn \cite{smoldyn} and Aboria \cite{aboria}, respectively. DBW would like to thank the EPSRC (Grant number EP/G03706X/1) for funding through a studentship at the Systems Biology programme of The University of Oxford's Doctoral Training Centre.
\end{acknowledgments}





%

\end{document}